\documentclass[prl,aps,floats, superscriptaddress,longbibliography, twocolumn]{revtex4-1}
\usepackage{amsmath}
\usepackage{amssymb}
\usepackage{graphics}
\usepackage{graphicx}
\usepackage{hyperref}
\usepackage{bm}

\newcommand{\ket}[1]{|#1\rangle}
\newcommand{\bra}[1]{\langle #1|}
\newcommand{\braket}[2]{\langle #1|#2\rangle}
\newcommand{\braOket}[3]{\langle #1|#2|#3\rangle}

\hypersetup{colorlinks,linkcolor=blue,urlcolor=blue,citecolor=blue}

\begin{document} 

\title{Variational Ansatz for an Abelian to non-Abelian Topological Phase Transition in $\nu=1/2+1/2$ Bilayers}
\date{\today}
\author{Valentin Cr\'epel}
\affiliation{Laboratoire de Physique de l'\'Ecole normale sup\'erieure, ENS, Universit\'e PSL, CNRS, Sorbonne Universit\'e, Universit\'e Paris-Diderot, Sorbonne Paris Cit\'e, Paris, France.}
\author{Benoit Estienne}
\affiliation{Sorbonne Universit\'{e}, CNRS, Laboratoire de   Physique Th\'{e}orique et Hautes \'{E}nergies, LPTHE, F-75005 Paris, France.}
\author{Nicolas Regnault}
\affiliation{Laboratoire de Physique de l'\'Ecole normale sup\'erieure, ENS, Universit\'e PSL, CNRS, Sorbonne Universit\'e, Universit\'e Paris-Diderot, Sorbonne Paris Cit\'e, Paris, France.}

\begin{abstract}
	We propose a one-parameter variational ansatz to describe the tunneling-driven Abelian to non-Abelian transition in bosonic $\nu=1/2+1/2$ fractional quantum Hall bilayers. This ansatz, based on exact matrix product states, captures the low-energy physics all along the transition and allows to probe its characteristic features. The transition is continuous, characterized by the decoupling of antisymmetric degrees of freedom. We futhermore determine the tunneling strength above which non-Abelian statistics should be observed experimentally. Finally, we propose to engineer the inter-layer tunneling to create an interface trapping a neutral chiral Majorana. We microscopically characterize such an interface using a slightly modified model wavefunction.
\end{abstract}

\maketitle

\paragraph{Introduction ---} Fractional Quantum Hall (FQH) systems are to date the most promising plateform to investigate phases of matter with intrinsic topological order~\cite{Gossard_HugeMagnetoTransport,Laughlin_AnsatzAndPlasma1,Wen_TopoOrderReview}. They host emergent quasiparticles which carry fractionalized quantum numbers~\cite{Etienne_FractionalCharge,Su_FractionalCharge} and obey anyonic statistics. Among them, non-Abelian FQH states are of fundamental importance with potential application to fault-tolerant topological quantum computations~\cite{KitaevKong_ModelGappedBoundary,Review_TopoQuantAndNonAbelian,Stern_ReviewNonAbelian}. Innovative experimental advances on heterostructure design~\cite{Heterostructure_DiracFermionVDW,Heterostructure_CompositeFermionTunable,Heterostructure_ChernInsulatorVDW,Heterostructure_Excitons} rapidly promoted these new setups as highly competitive for the study of strongly correlated quantum phases~\cite{MagicAngle_InsulatorHalfFilling,MagicAngle_Superconductivity}. At the same time, new layer-sensitive spectroscopic techniques are developed in wide and double quantum wells~\cite{LayerSpectroscopy_3D,LayerSpectroscopy_Eisenstein}, which allowed to probe the long discussed condensation of excitons~\cite{ExcitonBoseCondensate}. The inner (valley or layer) degree of freedom in these systems offers additional tunable parameters~\cite{Halperin_AllTunneling,PapicRegnaultGoerbig_331TunnelingFermiSea} and enriches the phase diagram~\cite{ValleyMatters_AbInitio,ValleyMatters_MacDonald,ValleyMatters_ExpMoSe}. This revives the interest in engineering non-Abelian topological phases from coupling internal degrees of freedom of multicomponent FQH systems~\cite{Halperin_SpinfulModel,HaldaneRezayi_SpinSinglet,ReadGreen_PairedStates,Fradkin_ChernSimonsPfaffian,Fradkin_GinzburgLandauNonAbelian} which are initially prepared in a well controlled Abelian state~\cite{Laughlin_AnsatzAndPlasma2,Jain_CompositeFermions}. 

Theoretically, the simplest construction of such an Abelian to non-Abelian transition starts with an Halperin 220 state~\cite{Halperin_Ansatz} at total filling fraction $\nu=1$. It describes two decoupled copies of the bosonic Laughlin 1/2 wavefunction (WF)~\cite{Laughlin_AnsatzAndPlasma1,Laughlin_AnsatzAndPlasma2}. Symmetrization of the two copies~\cite{Regnault_ProjectiveMearuementZk,Capelli_Projectiveconstruction,Cabra_AbelianToNonAbelianProjected,SuppMat} leads to the non-Abelian bosonic Pfaffian state~\cite{MooreRead_CFTCorrelatorModelState}. Symmetry gauging arguments~\cite{Barkeshli_OrbifoldLongPaper,BondersonBarkeshli_SymGauging,TeoFradkin_TwistLiquids} have shown that this procedure not only relates the two WFs, but it also produces the full non-Abelian Pfaffian topological order~\cite{WilczekNayak_NonAbelianMR} from the Abelian Halperin one~\cite{WenZee_Kmatrix}. It was argued~\cite{ReadGreen_PairedStates} that interlayer tunneling could physically drive such a symmetrization of the low energy degrees of freedom. This statement was put on firmer ground by considering the effect of tunneling  on the one-dimensional effective theory at the edge of the system~\cite{Sondhi_ChiralSineGordon,Kun_PhaseTransition331EdgeTheoryInterface}, while numerical studies have repeatedly confirmed that the tunneling-driven Abelian to non-Abelian transition occurs at a microscopic level in bilayer FQH systems~\cite{Sheng_BosonicBilayerTunnelingDMRG,ShengHaldane_Tunneling331MR,RegnaultGoerbig_BridgeAbToNonAb,PapicRegnaultGoerbig_331TunnelingFermiSea,Mong_CompetingOrdersFermionicBilayerDMRGandED,Jolicoeur_SpinlessBosons,Ueda_PhaseDiagBilayerBoson}. 

Model WFs have widely contributed to our understanding of correlated phases of quantum matter such as the BCS ansatz~\cite{Cooper_CooperPairs,BCS_Microscopic,BCS_Theory} and FQH model states~\cite{Laughlin_AnsatzAndPlasma1,Laughlin_AnsatzAndPlasma2,Halperin_Ansatz,Jain_CompositeFermions,MooreRead_CFTCorrelatorModelState}. Their physical relevance were soon corroborated by the finding of Hamiltonians for which they are the exact ground state: the Bogoliubov approach to superconductivity~\cite{Bogoliubov_FirstPaper,Bogoliubov_Review} and model $N$-body interactions for the FQHE~\cite{Kivelson_PseudoPotentials,Haldane_HierarchiesPseudoPot,HaldaneBernevig_ClusteringCondition}. Although the interactions stabilizing the FQH model states are not realistic~\cite{Haldane_PseudoPot}, the corresponding ground state WFs nonetheless capture the universal features of the experimentally observed phases such as quasiparticle charge and braiding statistics in the bulk and quasihole exponents on the gapless edge. The connection between FQH phases and the underlying topological order was considerably substantiated by Moore and Read in Ref.~\cite{MooreRead_CFTCorrelatorModelState}. They identified a large class of model WFs and their quasihole excitations with Conformal Field Theory (CFT) correlators from which the topological content of the phase may be read off (under the generalized screening assumption~\cite{Bonderson_PlasmaIsingHall}). It furthermore allows for an exact Matrix Product State (MPS) description of these strongly correlated phases of matter~\cite{Sierra_MPSfromCFT,ZaletelMong_ExactMPS,Regnault_ConstructionMPS}, allowing for large scale numerical study of their relevance and properties~\cite{Regnault_MPSCorrelationLength,Regnault_MPSBraiding}.

In this article, we propose a variational ansatz based on the CFT description in order to fully capture the low energy physics of a bosonic FQH bilayer with \emph{arbitrary} interlayer tunneling. The MPS description allows us to observe a continuous phase transition driven by the decoupling of antisymmetric degrees of freedom. We also determine the precise range of tunneling where a non-Abelian order fully develops.

\paragraph{Microscopic Model ---}

We study a two-component bosonic system at total filling $\nu=1$ in the FQH regime. Bosons populate the lowest band assumed to be a Chern band with $\mathcal{C}=1$~\cite{Thouless_ChernNumber}, nearly flat and separated from the other ones by a large gap. For simplicity, we focus on the continuum limit where a large uniform magnetic field (possibly artificial~\cite{Dalibard_ArtificialGauge,GoldmanNascimbene_ReviewArtificialGauge}) produces exactly flat Landau levels. Temperature and interaction strength are supposed to be small enough to avoid Landau level mixing, allowing projection onto the Lowest Landau Level (LLL). The layer degree of freedom forms a pseudo spin-1/2 with $z$-components $\sigma \in \{\uparrow, \downarrow\}$. Although we will use the vocabulary of bilayer systems, our discussion extends to any other type of two-dimensional internal degree of freedom such as spin, hyperfine states of an atom, and so on. As kinetic energy is frozen, the interactions projected to the LLL dominate and produce strong correlations between particles. We assume density-density interactions within each layers
\begin{equation} \label{eq:InteractingHamiltonian}
\mathcal{H}_{\rm int} = J \int {\rm d}^2 z \sum_{\sigma \in \{\uparrow, \downarrow\}} : \rho_\sigma (z) \rho_\sigma (z) : \, ,
\end{equation} where we have defined $\rho_\sigma (z)$ the LLL projected density in layer $\sigma$ at position $z$. The interaction strength $J$ is set such that the two-body energy-scale is one. The densest ground state of Eq.~\ref{eq:InteractingHamiltonian} is the Halperin 220 state~\cite{Kivelson_PseudoPotentials,Thomale_PseudoPotentials}. It has a four-fold topological degeneracy on the torus, corresponding to the four anyons of the underlying Abelian topological order~\cite{WenZee_Kmatrix}. Under the global layer permutation $\mathcal{P}_z$ (swapping $\uparrow$ and $\downarrow$) which commutes with $\mathcal{H}_{\rm int}$, three of them are even while the last one is odd~\cite{Regnault_BilayerBosonicPhaseDiagram}. Coupling of the two layers is achieved by introducing tunneling as $\mathcal{H}_{\rm tun} = -J t S_x$ .
It acts as a Zeeman term coupling to the $x$-components of the total spin $\mathbf{S}$ and splits the Halperin 220 ground state manifold into even and odd sectors. With increasing tunneling strengths, the odd-parity state crosses the gap and ultimately merges into the many-body spectrum continuum~\cite{Sheng_BosonicBilayerTunnelingDMRG}. The remaining three-fold ground state degeneracy is a signal of the non-Abelian Pfaffian topological order~\cite{MooreRead_CFTCorrelatorModelState}, which we can understand as follows. Tunneling splits the LLL into two sub-bands, gathering respectively single-particle WFs with layer index in the two pseudo-spin $x$-eigenstates, which we denote as even $e = (\uparrow+\downarrow)/\sqrt{2}$ and odd $o = (\uparrow-\downarrow)/\sqrt{2}$. At large tunneling $t\gg 1$, the system behaves as an effective single-component FQH system fully polarized in $e$, due to the large sub-band Zeeman gap, with delta interactions. We rely on numerical studies to certify that its ground state is well approximated by the non-Abelian Pfaffian state~\cite{Regnault_BilayerBosonicPhaseDiagram,RegnaultGoerbig_BridgeAbToNonAb}.

\paragraph{Model State ---}

Motivated by the solution of a specific model of spin-triplet pairing in $p$-wave superconductors displaying a similar tunneling-driven Abelian to non-Abelian transition~\cite{SuppMat,ReadRezayi_RouteToNonAbelianZeroModes,ReadGreen_PairedStates}, we propose the following one-parameter ansatz for the low energy theory of $\mathcal{H}_{\rm int} + \mathcal{H}_{\rm tun}$:
\begin{widetext} \begin{equation} \label{eq:VariationalAnsatzTheta}
	\ket{\Psi_\theta \rangle } = \braOket{0}{ \mathcal{O}_{\rm Bkg} \, \exp\left[ \int {\rm d}^2z \left( \cos\theta \, \mathcal{V}^e(z) \otimes c_e^\dagger(z) + \sin\theta \, \mathcal{V}^o(z) \otimes c_o^\dagger(z) \right) \right]}{0} \otimes \ket{\Omega\rangle} \, , \, \theta \in [0, \pi/4] \, . 
	\end{equation} \end{widetext} Here, $c_{e/o}^\dagger(z)$ creates a boson in the spin component $e/o$ at position $z$ above the Fock vacuum $\ket{\Omega\rangle}$. $\ket{0}$ is the CFT vacuum state. The CFT uses two chiral bosonic field $\phi^s$ and $\phi^c$~\cite{YellowBook} to encode the spin and charge degrees of freedom which decouple on the edge of the system~\cite{Voit_SpinCharge,Giamarchi_1Dphysics}. $\phi^c$ carries a U(1)-charge associated with particle number. The neutralizing background charge $\mathcal{O}_{\rm Bkg}$ fixes the overall number of particles and reproduces the Gaussian factors of the LLL~\cite{MooreRead_CFTCorrelatorModelState,Regnault_MPSnonAbelianQH}. Spin excitations are described by a Dirac fermion $\Psi^\dagger = :e^{i\phi^s}:$~\cite{Sondhi_ChiralSineGordon,Cabra_AbelianToNonAbelianProjected}. Its real and imaginary parts are related to the spin in the $x$-direction. The operators associated with the bosonic creation operators in Eq.~\ref{eq:VariationalAnsatzTheta} are
\begin{equation}\label{eq:Electronicoperators}
\mathcal{V}^e =  :\cos \left( \phi^s \right) e^{i \phi^c}: \quad , \quad \mathcal{V}^o  = i:\sin \left( \phi^s \right) e^{i \phi^c}: \, .
\end{equation} Our ansatz smoothly interpolates between the Pfaffian and the Halperin 220 state which are exactly reproduced respectively for $\theta=0$ and $\theta=\pi/4$~\cite{SuppMat}. As $\theta$ goes to zero, the total spin of the variational ansatz polarizes in the $e$ component, which drives the transition discussed above.

\begin{figure}
	\centering
	\includegraphics[width=\columnwidth]{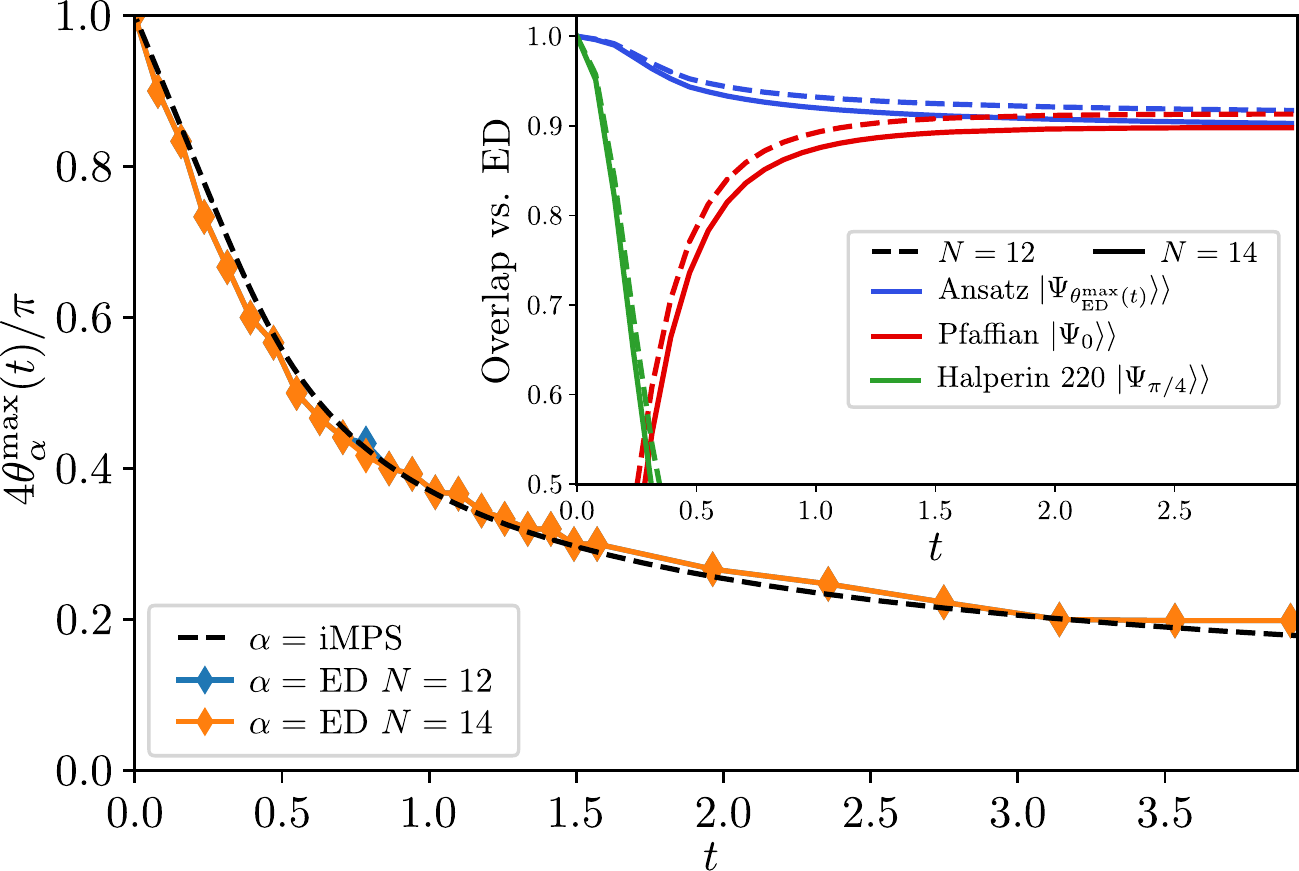}
	\caption{Comparison of our ansatz with the lowest momentum ED ground state on a cylinder of perimeter $L=8\ell_B$. Main: The variational parameter $\theta_{\rm ED}^{\rm max}(t)$ maximizing the overlap for $N=12$ and $N=14$ particles as a function of tunneling (diamonds). It agrees extremely well with $\theta_{\rm iMPS}^{\rm max}(t)$ determined independently by energy minimization on an infinite cylinder (dashed line). Inset: Best overlap with our variational ansatz (blue) for two system sizes as a function of the tunneling strength $t$. For comparison, the overlap with the Pfaffian (red) and Halperin 220 (green) states are also provided. Our one-parameter variational ansatz correctly captures the low energy physics of the system over the whole range of $t$, even when the other two model states perform poorly.}
	\label{Fig:EnergyMinimizationEDvsInfinite}
\end{figure}

Being written as a CFT correlator, our ansatz can be brought to an MPS form. All numerical calculations are performed on a cylinder of perimeter $L$ (measured in units of magnetic length $\ell_B$), the efficient geometry for the FQH MPS formulation~\cite{ZaletelMong_ExactMPS,Zaletel_DMRG_Hall,Zaletel_DMRG_Multicomponent}. We first computed the overlap between the ED ground state of lowest momentum for several tunneling strength $t$. The best variational parameter $\theta_{\rm ED}^{\rm max}(t)$ is depicted in Fig.~\ref{Fig:EnergyMinimizationEDvsInfinite}. Since both the Halperin 220 and the Pfaffian states belong to the optimization set, it is not surprising that our ansatz performs better than these two model states (also displayed in Fig.~\ref{Fig:EnergyMinimizationEDvsInfinite}). However in the transition region where they both fail to capture the low energy physics, our ansatz remains a very good approximation of the ground state. Hence, the theoretical ingredients used to build Eq.~\ref{eq:VariationalAnsatzTheta} seem to faithfully account for the low energy physics of the model for any tunneling. To confirm these high overlaps are not mere artifacts of the optimization procedure, we can fix the variational parameter \emph{independently} by minimizing the energy per orbital on an infinite cylinder. We developed a method to evaluate the interaction $E(\theta) = \braOket{\langle \Psi_\theta}{\mathcal{H}_{\rm int}}{\Psi_\theta \rangle}$ and tunneling $T(\theta) = \braOket{\langle \Psi_\theta}{\mathcal{H}_{\rm tun}}{\Psi_\theta \rangle}$ energies which bypasses the use of matrix product operators~\cite{SuppMat}. In essence, we have adapted known results about continuous MPS~\cite{Cirac_CalculusContinuousMPS} to our ansatz $\ket{\Psi_\theta \rangle}$, seeing it as a Hamiltonian time evolution of a 1D CFT~\cite{Regnault_MPSnonAbelianQH}. The energy minimization procedure fixes the variational parameter $\theta_{\rm iMPS}^{\rm max} (t)$, as depicted in Fig.~\ref{Fig:EnergyMinimizationEDvsInfinite}. We find an excellent agreement between the numerically extracted $\theta_{\rm iMPS}^{\rm max} (t)$ and our previous finite size results $\theta_{\rm ED}^{\rm max} (t)$. This provides another stringent test of the physical relevance of our ansatz.

\paragraph{Continuous Transition ---} 

\begin{figure}
	\centering
	\includegraphics[width=\columnwidth]{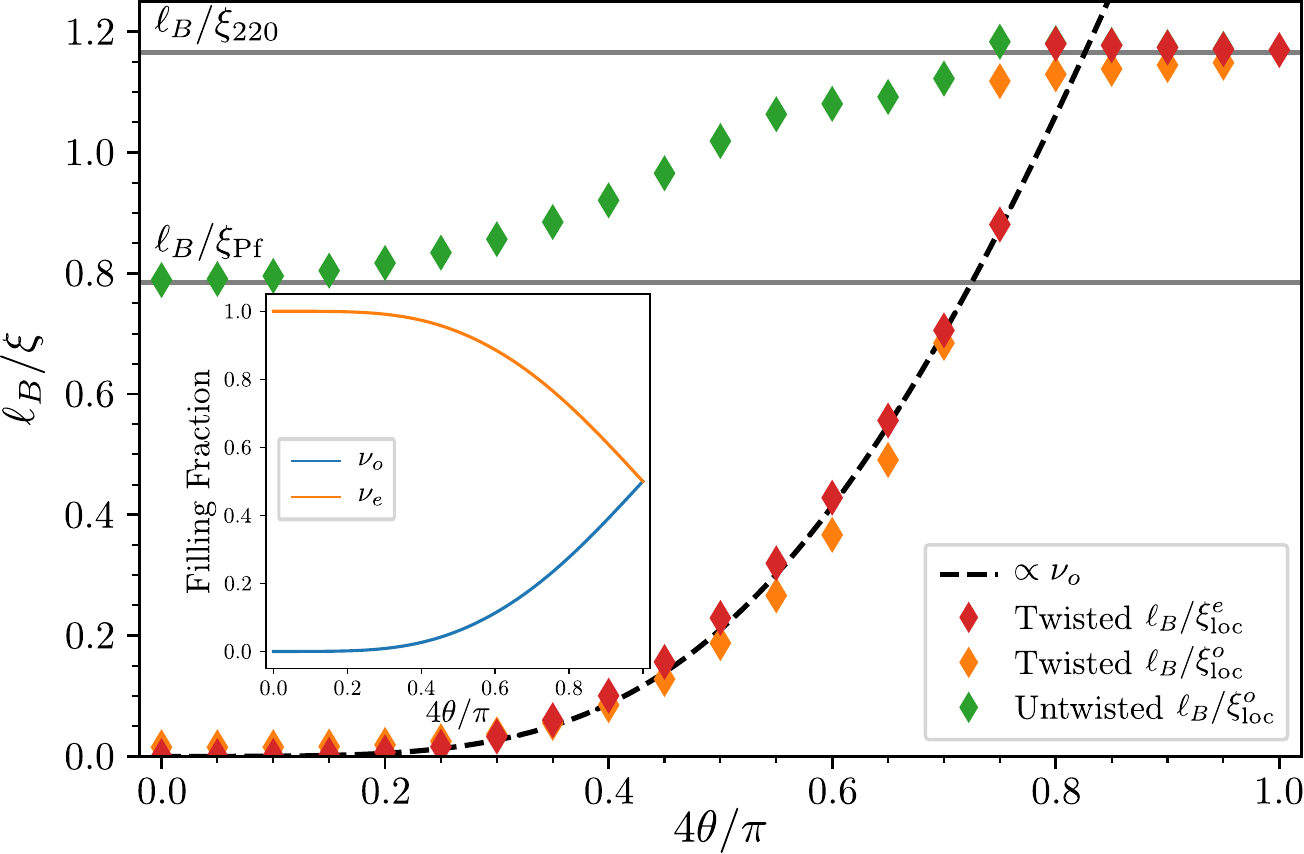}
	\caption{Main: Inverse of the even and odd correlation lengths as a function of the variational parameter. The even correlation length $\xi_{\rm loc}^e$ remains finite and interpolates between the Halperin 220 $\xi_{\rm 220}$ and Pfaffian $\xi_{\rm Pf}$ values. The odd correlation length, computed in the twisted and untwisted sectors (see below), diverges as $\ell_B/\xi_{\rm loc}^o\propto \nu_o$ when $\theta \to 0$ (sse text). For $4\theta/\pi > 0.7$, a level crossing in the spectrum prevents us to follow the corresponding transfer matrix eigenstate. Inset: Filling fraction $\nu_o$ in blue (resp. $\nu_e$ in orange) of the odd (resp. even) LLL band of Eq.~\ref{eq:VariationalAnsatzTheta}. }
	\label{Fig:CorrelLength}
\end{figure}

For a state to exhibit an intrinsic topological order, the ground state manifold has to be immune to local perturbations. The system should be gapped and as a consequence, correlation function of local observables decay exponentially with distance. This is usually dubbed as screening~\cite{Bonderson_PlasmaIsingHall} in the FQH language due to the plasma analogy~\cite{Laughlin_AnsatzAndPlasma1}. We investigate the screening properties of the system thanks to the MPS description of our ansatz. In the MPS formalism, a local excitation such as $c_e (z)\ket{\Psi_\theta \rangle}$ is translated into the insertion of $\mathcal{V}^e$ (see Eq.~\ref{eq:VariationalAnsatzTheta}) which couples to an excited state of the transfer matrix. The correlation length governing the decay of $\braOket{\langle \Psi_\theta}{c_e^\dagger (z) c_e (w)}{\Psi_\theta \rangle}$ is then related to the corresponding eigenvalue of the transfer matrix~\cite{GapTransferMatrixCorrelation,Regnault_MPSCorrelationLength}. While bulk excitations generically couple to the first excited state of the transfer matrix, the $\mathcal{P}_z$ symmetry, which translates into $\phi^s \to -\phi^s$ in the CFT, imposes further selection rules. Symmetric (resp. anti-symmetric) excitations under layer inversion couple to even (resp. odd) excited state under $\mathcal{P}_z$. We extracted the corresponding even $\xi_{\rm loc}^e$ and odd $\xi_{\rm loc}^o$ correlation lengths as a function of $\theta$, as shown in Fig.~\ref{Fig:CorrelLength}. The correlation length $\xi_{\rm loc}^e$ is finite for all $\theta$ and smoothly interpolates between the values of Refs.~\cite{Regnault_MPSHalperinStates,Regnault_MPSCorrelationLength} for the Halperin 220 and one-component bosonic Pfaffian states. We observe that antisymmetric excitations are not equally well screened and that $\xi_{\rm loc}^o$ diverges when $\theta$ goes to zero. We relate these features to the depletion of the $o$ band. Notice indeed that $\mathcal{V}^o$ insertions in Eq.~\ref{eq:VariationalAnsatzTheta} are only screened by $o$ electrons with which they have non trivial fusion rules. Linear response theory predicts that $\xi_{\rm loc}^o$ is inversely proportional to $\nu_o$ when the filling factor of the $o$ band goes to zero. We find that this law accurately accounts for the computed $\xi_{\rm loc}^o$, as depicted on Fig.~\ref{Fig:CorrelLength}. We shall now argue that this critical behavior at small $\theta$ remains unnoticed. Take $\theta=0$, the system has no $o$ particles (see Eq.~\ref{eq:VariationalAnsatzTheta}) and only couples to the symmetric part of local observables, for instance: \begin{equation} \label{eq:DecouplingEvenOdd}
\braOket{\langle \Psi_0}{c_\uparrow^\dagger (z) c_\uparrow (w)}{\Psi_0 \rangle} = \braOket{\langle \Psi_0}{c_e^\dagger (z) c_e (w)}{\Psi_0 \rangle} \, .
\end{equation} This creates some redundancies in the MPS description since the antisymmetric part of the $\phi^s$ field is conserved. The corresponding degrees of freedom on the MPS boundary condition lead to exact degeneracies in the transfer matrix spectrum and explain the divergence of $\xi_{\rm loc}^o$ at $\theta=0$. Note however that all these different MPS boundary conditions are completely transparent to the $e$ particles and produce the \emph{same} physical state. For $\theta \lesssim \pi/20$, the bands are only weakly coupled as shown in Fig.~\ref{Fig:CorrelLength}. The system is almost polarized $\nu_o \ll \nu_e$ and exhibit a clear scale separation $\xi_{\rm loc}^e \ll \xi_{\rm loc}^o$. The previous degeneracies are slightly lifted and excitations in the empty $o$ band form a gapless branch above the ground state. However, $o$ degrees of freedom only become noticeable at very large scales $\sim \xi_{\rm loc}^o$. For instance, the first correction to Eq.~\ref{eq:DecouplingEvenOdd} is of order $\sqrt{\nu_o} e^{-|z-w|/\xi_{\rm loc}^o}$, small in magnitude unless the even excitation is fully screened. Low $\nu_o$ population makes these excitations transparent on the scale over which the correlation builds up in the $e$ band, such that the ground state properties are locally that of the $\theta=0$ point. 

Thanks to the MPS ansatz, the same analysis can be carried for anyon-anyon correlation functions which are described by non-local operators~\cite{Schuch_PepsBilayer,Schuch_PEPSstudyAnyonCondensation}. We observe a similar decoupling of the antisymmetric degrees of freedom, which do not interfere with the low energy physics of the $e$ electrons in the small $\theta$ limit~\cite{SuppMat}.

\paragraph{Non-Abelian Properties ---} At $\theta = 0$, we saw that the odd excitations decouple from the low-energy degrees of freedom of the system. Splitting the Halperin 220 anyonic excitations into even and odd parts under $\mathcal{P}_z$, we are left with two deconfined anyons $\psi$ and $\sigma = : \cos\left(\phi^s/2\right) e^{\frac{i}{2} \phi^c}:$ obeying the fusion rules:
\begin{equation} \label{eq:MR_Fusion}
\psi \times \psi = \mathbf{1} \, , \quad \psi \times \sigma = \sigma \, , \quad \sigma \times \sigma = \mathbf{1} + \psi \, .
\end{equation} 
They define an $SU(2)_2$ algebra~\cite{BaisSlingerland_CosetAnyonCondensation}, which characterizes the Pfaffian topological order~\cite{MooreRead_CFTCorrelatorModelState}. The previous discussion about the correlation lengths establishes that the same non-Abelian theory extends above this special point, here for $\theta \lesssim \pi/20$. The Pfaffian physics can be probed by braiding anyons kept by a few $\xi_{\rm loc}^e$ apart and the observed separation of scale ensures that odd contributions do not spoil the results. Going from the Halperin 220 order to Eq.~\ref{eq:MR_Fusion} can be interpreted as gauging the symmetry $\mathcal{P}_z$~\cite{BondersonBarkeshli_SymGauging,RyuTeo_OrbifoldToGauging}, which is indeed locally satisfied when the system is fully polarized in $e$, followed by the deconfinement of the symmetry defects~\cite{TeoFradkin_TwistLiquids} which we couldn't probe directly though we believe it to be related to the $\xi_{\rm loc}^o$ divergence. Restoring the Abelian order from Eq.~\ref{eq:MR_Fusion} can be viewed as condensing of the boson $i \partial \phi^s$~\cite{Barkeshli_OrbifoldLongPaper}.

\begin{figure}
	\centering
	\includegraphics[width=\columnwidth]{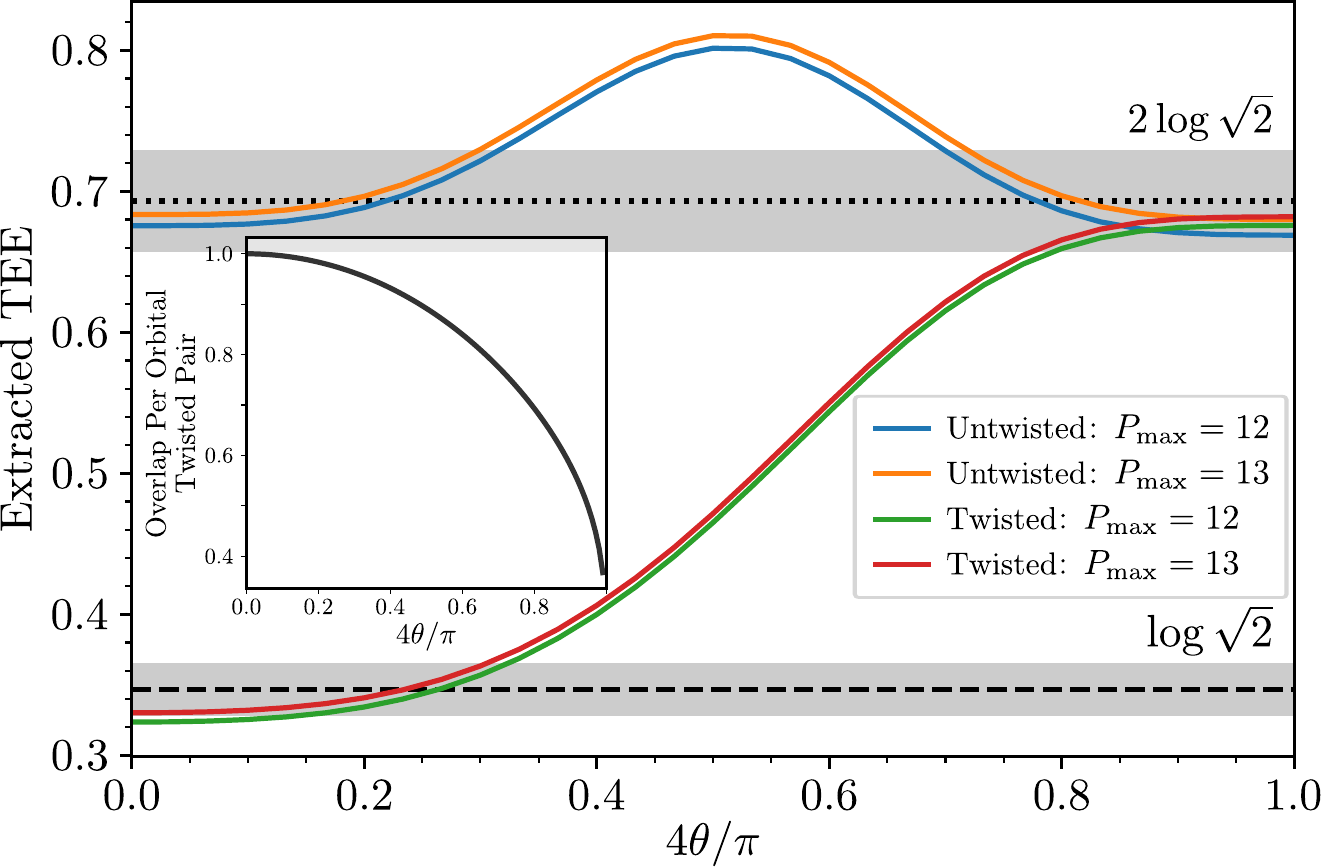}
	\caption{Main: Extraction of the TEE at $L=10\ell_B$ in the twisted and untwisted sectors using the method of Ref.~\cite{Regnault_MPSHalperinStates}. Dotted lines indicate theoretical predictions and gray shaded areas around them represent a 5\% deviation. While we are still off these prediction by a few percents in the Halperin 220 and Pfaffian phases, increasing the MPS bond dimension measured by $P_{\rm max}$~\cite{SuppMat} brings our results closer to the expected values. Inset: Overlap per orbital between the two physical states built with MPS boundary conditions in the two sectors of the twisted pair for a cylinder of perimeter $L=10\ell_B$. As $\theta \to 0$, the overlap reaches one showing that they describe the same physical state.}
	\label{Fig:TEE_ComputaedAsTheta}
\end{figure}

To test the non-Abelian character at small $\theta$ and put bound on the Pfaffian regime in the phase diagram, we evaluated the constant correction to the entanglement area law, the Topological Entanglement Entropy (TEE)~\cite{KitaevPreskill_TEE,LevinWen_TEE}, which is known to characterize the topological order. We rely on the ansatz structure, that we shortly sum up. The CFT Hilbert space splits into four sectors, stable under the action of the bosonic operators $\mathcal{V}^{e/o}$. We can group them into two pairs that we denote as twisted or untwisted, depending on the Dirac fermion boundary condition when it winds around the cylinder. We can thus build four physical states on the infinite cylinder differing by the MPS boundary condition. For the Halperin 220 state, they correspond to the four degenerate ground states and the odd sector is in the twisted pair. At the Pfaffian point, the two twisted sectors produce the same physical state, their overlap is one as shown in Fig.~\ref{Fig:TEE_ComputaedAsTheta} due to a redundancy in the CFT Hilbert space~\cite{SuppMat}. The ground state is thus threefold degenerate as expected. 

We computed the real-space entanglement entropy $S(\theta)$ of Eq.~\ref{eq:VariationalAnsatzTheta} in all sectors, for a cut preserving the cylinder rotational symmetry~\cite{RSESdubail,RSESsterdyniak,RSESRodriguezSimonSlingerland}. The TEE is then extracted by finite difference with respect to the cylinder perimeter, as in Refs.~\cite{Regnault_ConstructionMPS,OurNatComms_H2L2,OurNatComms_H3L3}. The numerical results for one twisted and one untwisted sector are depicted in Fig.~\ref{Fig:TEE_ComputaedAsTheta}. Identical results were found for both members of the (un)twisted pair, which are related by a center of mass translation corresponding to a
shift of the U(1)-charge at the MPS boundary. When $\theta \simeq \pi/4$ all four sectors have the same TEE $\gamma \simeq 2 \log \sqrt{2}$ matching the prediction for the Abelian Halperin 220 topological order~\cite{WenZee_Kmatrix,Hansson_ReviewCFTforFQHE}. For $\theta \lesssim \pi/20$, we find $\gamma_\mathbb{I} = \gamma_\psi \simeq \log \sqrt{4}$ in the untwisted sectors, and $\gamma_\sigma \simeq \log \sqrt{2}$ in the twisted sector. The different sectors having distinct TEE is an indication of the non-Abelian nature of the phase. The extracted TEEs agree with the quantum dimensions of the $\mathbb{I}$, $\psi$ and $\sigma$ anyons~\cite{BaisSlingerland_CosetAnyonCondensation}. These results bolster the physical picture depicted above. They also provide a quantitative range $\theta$ and hence in $t$ over which the system exhibit a true Abelian ($\theta \in [\pi/5 ,\pi/4]$, $t \in [0 ,0.2162(1)]$) or non-Abelian ($\theta \in [0,\pi/20]$, $t \in [2.74(7) , \infty[$) order.

\paragraph{Trapping a Majorana ---} It is argued that a one-dimensional neutral chiral Majorana fermion is trapped at the interface between a Halperin 220 phase and a Pfaffian phase~\cite{BaisSlingerland_CosetAnyonCondensation,Kun_PhaseTransition331EdgeTheoryInterface}. Our study provides a simple protocol to realize such a setup microscopically, varying the tunneling parameter spatially. Let $t(x)$ vary smoothly along the cylinder axis such that the Halperin 220 and Pfaffian phases are stabilized on either sides of the interface, where the interaction Eq.~\ref{eq:InteractingHamiltonian} mixes their counter-propagating edge modes. Assuming that all interface degrees of freedom present on both sides gap out~\cite{BaisSlingerland_CosetAnyonCondensation,Kun_PhaseTransition331EdgeTheoryInterface}, only the antisymmetric edge excitations remain. They are described by a chiral Majorana fermion $\psi^I$, the imaginary part of $\Psi$.

\begin{figure}
	\centering
	\includegraphics[width=\columnwidth]{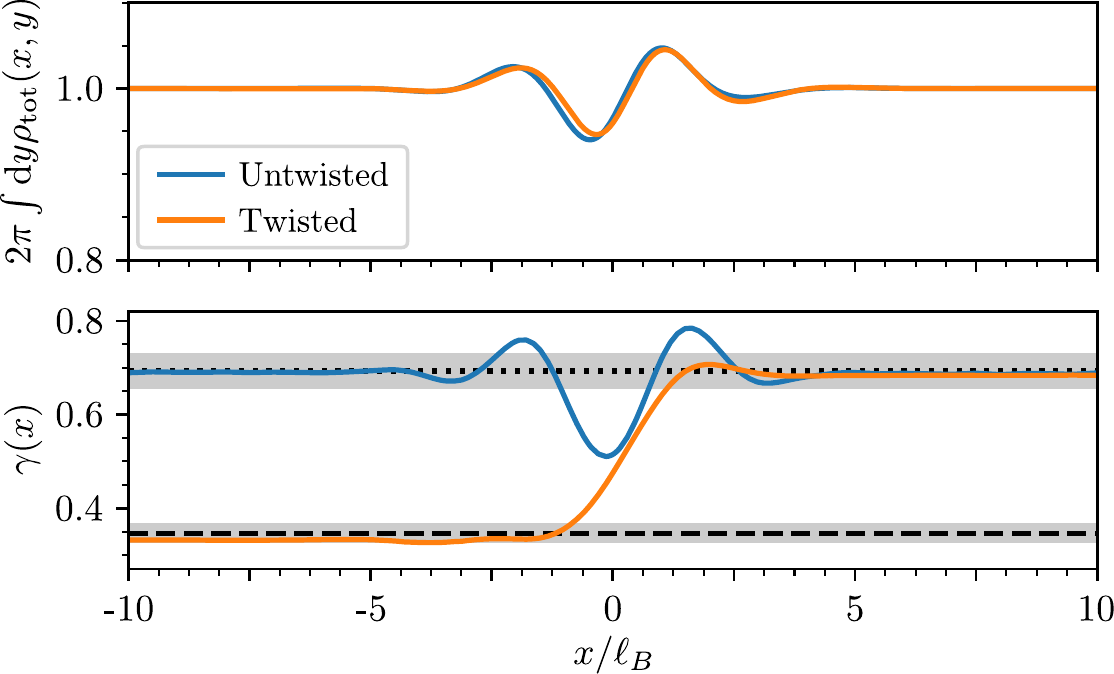}
	\caption{Top: Total density integrated over the cylinder perimeter across the interface. It is featureless, but for small ripples certainly due to the sharpness of our interface ansatz (see Refs.~\cite{OurNatComms_H2L2,OurNatComms_H3L3}). We checked that no charge accumulates at the interface, even for excitations of $\psi^I$, showing that the interface effective theory is neutral. Bottom: Extracted TEE in the twisted and untwisted sectors at different position across the interface. We recover the values characterizing the Pfaffian and Halperin 220 phases away from the interface $|x| \gg \ell_B$, showing the relevance of the interface MPS WF.}
	\label{Fig:Interface_Charac}
\end{figure}

We can numerically probe such an interface between distinct topological orders within the MPS formalism as first described in Refs.~\cite{OurNatComms_H2L2,OurNatComms_H3L3}. A model WF describing a sharp interface at $x=0$ is built by using the Halperin (resp. Pfaffian) MPS matrices for LLL orbitals centered at $x>0$ (resp. $x<0$). This interface model WF was shown to capture all universal as well as some microscopic features of the interface~~\cite{OurNatComms_H2L2,OurNatComms_H3L3}. We first extract the TEE $\gamma(x)$ at different position $x$, the results are depicted on Fig.~\ref{Fig:Interface_Charac}b. Deep in the two phases, we retrieve the expected TEE for the Halperin 220 and Pfaffian phases, which proves that our interface model WF indeed interpolates between the two topological orders. Next, we observe that, besides small ripples, the density is completely featureless across the interface (see Fig.~\ref{Fig:Interface_Charac}a). We checked this to also be true for excited state of $\psi^I$, which shows that the interface edge mode is neutral. To further characterize the interface low-energy effective theory, we extracted its chiral anomaly (or central charge) $c$ following the methods developed in Refs.~\cite{OurNatComms_H2L2,OurNatComms_H3L3}. We find $c \simeq 1/2$~\cite{SuppMat}, for which the only unitary minimal model is that of a free Majorana fermion~\cite{YellowBook}.

\paragraph{Conclusion ---} In this letter, we extended the CFT approach to bulk FQH WFs to describe a continuous phase transition between distinct topological orders. Besides the two fixed points on either sides of the transition, it accurately describes the low-energy physics of FQH bosonic bilayers at arbitrary tunneling. We used it to observe the characteristic features of the Abelian to non-Abelian transition and could put bounds on each domains. Finally, we showed that a neutral Majorana fermion could be trapped at an interface engineered by spatially varying the tunneling strength.

\paragraph*{Acknowledgement} 

We thank Y. Fuji, J. Slingerland and P. Lecheminant for enlightening discussions. VC, BE and NR were supported by the grant ANR TNSTRONG No. ANR-16-CE30-0025 and ANR TopO No. ANR-17-CE30-0013-01.

\bibliography{mr220transition}

\onecolumngrid
\appendix
\pagebreak
\begin{center}
	\textbf{\Large Supplemental Materials: Variational Ansatz for an Abelian to non-Abelian Topological Phase Transition in $\nu=1/2+1/2$ Bilayers}
\end{center}

\setcounter{equation}{0}
\setcounter{figure}{0}
\setcounter{table}{0}
\setcounter{page}{1}

\makeatletter
\renewcommand{\theequation}{E\arabic{equation}}
\renewcommand{\thefigure}{F\arabic{figure}}
\makeatother

\section{A. Triplet Pairing in $p$-wave Superfluids}

We give an overview of the model of $p$-wave pairing in a spinful superfluid system introduced in Refs.~\cite{ReadRezayi_RouteToNonAbelianZeroModes,ReadGreen_PairedStates}. It exhibits a tunneling-driven Abelian to non-Abelian phase transition similar to the one studied in the main text. For a general discussion, we refer the reader to the two previous references. Here, we focus on a certain parameter regime for which the transition is captured by an exact ground state that may be casted in a form similar to that of our variational ansatz (see main text). This provides another indication that our ansatz captures the relevant physics of such a transition, and justifies its form. It also allows for a better understanding of the features observed in the article such as the decoupling of odd degrees of freedom at large tunneling. Still, the special high symmetry line targeted in the phase diagram is highly non-generic and it overlooks many important microscopic details such as any UV regularization of the pairing function or the self-consistent superfluid gap equation. These potential caveats do not plague the model described in the main text.

We consider a He${}^3$ superfluid system exhibiting complex $p$-wave pairing~\cite{SuperfluidPhaseHelium3}, subject to a magnetic field along the quantization axis which we assume in the $x$-direction. Within BCS theory~\cite{BCS_Theory}, the Hamiltonian reads in momentum space: \begin{equation} \label{eq:BCS_Hamilto}
\mathcal{H}_{A-B} = \sum_k \zeta_k \left( c_{k,e}^\dagger c_{k,e} + c_{k,o}^\dagger c_{k,o} \right) - t \left( c_{k,e}^\dagger c_{k,e} - c_{k,o}^\dagger c_{k,o} \right) + \frac{1}{2} \sum_{\tau \in \{e,o\}} \left( \Delta_{k,\tau}^*  c_{k,\tau}^\dagger c_{-k,\tau}^\dagger + \Delta_{k,\tau}  c_{-k,\tau} c_{k,\tau} \right) \, .
\end{equation} The first term is the kinetic energy of the particles which have the same mass and chemical potential, \textit{i.e.} $\zeta_k = \frac{|k|^2}{2m}-\mu$. The second term is the Zeeman splitting due to the magnetic field along the $x$-axis. The last term corresponds to interactions treated within the BCS formalism. For complex $p$-wave pairing $\Delta_{k,\tau}$ carries angular momentum $\ell=-1$, leading to $\Delta_{k,\tau} \simeq \Delta_\tau (k_x-ik_y)$ up to some small distance regularization. Besides a frozen kinetic energy and a non-perturbative treatment of interactions, the model studied in the main text shares the same physical ingredients. We can view Eq.~\ref{eq:BCS_Hamilto} as two decoupled $p$-wave superfluids which are respectively even ($e$) and odd ($o$) under layer exchange symmetry (here interpreted as a spin exchange along $z$). Tunneling energetically favors $e$ particles, and the polarization along this spin component is seen as a chemical potential bias: $\mu_e = \mu + t$ and $\mu_o = \mu - t$. When tunneling reaches $t=\mu$, the $o$ subsystem undergoes a transition from a weak to a strong pairing phase, while $e$ particles remain in a weak pairing phase. The ground state is expected to host non-Abelian excitations after the transition, unaffected by the strongly-paired $o$ particles which decouple from the ground state~\cite{ReadGreen_PairedStates}.

We choose to cross this transition with $t\in [0,\mu]$ and fix the superfluid gap at $\Delta_{k,\tau} = \sqrt{\frac{2\mu_\tau}{m}} (k_x - k_y)$ for all momenta $k$ and $\tau \in \{e,o\}$. This allows to rewrite the BCS ground state as~\cite{DubailRead_ShortEE,RezayiDubail_InnerProduct}: \begin{equation} \label{eq:BCS_WFsAnsatz}
\ket{\Psi_{\rm BCS} \rangle} = \left\langle \exp\left[ \lambda \int {\rm d}^2 z \left( \cos \theta \psi^R(z) \otimes c_e^\dagger(z) + \sin \theta \psi^I(z) \otimes c_o^\dagger(z) \right) \right] \right\rangle \ket{\Omega\rangle} \, .
\end{equation} Here, we have separated the Dirac fermion $\Psi = (\psi^R - i\psi^I)/\sqrt{2}$ into its real and imaginary part which are both Majorana fields. The irrelevant constant $\lambda = 2 \sqrt{m \mu}$ could be absorbed in the WF normalization. Finally, we find that the angle $\theta$ satisfies: \begin{equation}
\theta = \arctan \left( \sqrt{ \frac{1-t/\mu}{1+t/\mu} } \right) \, ,
\end{equation} which interpolates between $\theta = \pi/4$ and $\theta =0$ when the tunneling is increased from zero to $\mu$ and triggers the weak-to-strong pairing transition of the $o$ particles. At the weak-to-strong pairing transition of the $o$ subsystem ($\theta=0$), there are no $o$ pairs in the ground state as in the main text. For $\theta$ finite but small, we find that local observables only couple to $e$ particles, up to corrections of order $\sin \theta \simeq \theta$ (two-point spin-$\uparrow$ correlators, pairing function, ...). The separation of scale observed in the main article can be understood by looking at the size of $oo$ pairs which diverging as $|1-t/\mu|^{-1/2}$ near the transition point in the strong coupling phase (which is not described by Eq.~\ref{eq:BCS_WFsAnsatz}).

\section{B. Derivation of the MPS}
\subsection{B.1. Overview of the General Construction}

We briefly recall how to obtain exact MPS for bosonic model states written as conformal blocks~\cite{ZaletelMong_ExactMPS,Regnault_ConstructionMPS}. For clarity, we exemplify our construction for a single component FQH state, though the same derivation holds for two-components model WFs, as shown in Ref.~\cite{Regnault_MPSHalperinStates}. We consider the flat cylinder geometry with perimeter $L$, \textit{i.e.} we assume periodic boundary condition along the $y$-axis. This geometry can be conformally mapped to the plane using $z = \exp(\gamma w)$ with $\gamma = 2\pi/L$ and $w=x+iy$ the cylinder coordinate. Because of this mapping, we shall use interchangeably $w$ or $z$ in the following discussion. A single particle basis spanning the Lowest Landau Level (LLL) reads in the Landau gauge
\begin{equation} \label{eq:LLLOrbitalCylinder}
\phi_j(w)= \mathcal{N}_j z^j e^{-\frac{x^2}{2\ell_B^2}} \quad \text{with: } \, \mathcal{N}_j = \frac{e^{-(\gamma \ell_B j)^2/2}}{\sqrt{L \ell_B \sqrt{\pi}}} \, .
\end{equation}  The one body states are labeled by the momentum along the cylinder perimeter $k_j = \gamma j$, with $j\in\mathbb{Z}$ when no magnetic flux thread the cylinder, which also determines their center $x_j = k_j \ell_B^2$ along the cylinder axis. From now on we set $\ell_B=1$. We note $c^\dagger(w)$ the bosonic creation operator at position $w$ and $c_j^\dagger =\int {\rm d}^2w \,  \phi_j(w) c^\dagger (w)$ the creation operator of a particle in the $j$-th orbital.

On the CFT side, $\mathcal{V}(w)$ is a primary of a rational CFT verifying the bosonic commutation relations $\mathcal{V}(w_1)\mathcal{V}(w_2) =  \mathcal{V}(w_2) \mathcal{V}(w_1)$ and whose OPE with itself only involves regular terms $\mathcal{V}(w_1)\mathcal{V}(w_2) \sim_{\rm \scriptscriptstyle  OPE} \mathcal{O}(1)$. For the model WF constructed over $\mathcal{V}(w)$, those conditions ensure that bosons braid trivially with respect to one another and have the correct commutation relations. $\mathcal{V}(w)$ is usually called the electronic operator in the literature~\cite{MooreRead_CFTCorrelatorModelState,Regnault_ConstructionMPS}, even for bosonic states. We will thus stick to this convention. In second quantization, the model state built out of the primary operator $\mathcal{V}$ can be written as~\cite{ZaletelMong_ExactMPS,RezayiDubail_InnerProduct}: 
\begin{equation} \label{eq:GenericMpsSecondQuant}
\ket{\Psi\rangle} = \braOket{0}{\mathcal{O}_{\rm Bkg} \, \exp\left[ \int {\rm d}^2w \, \mathcal{V} (w) \otimes c^\dagger(w) \right] }{0}\otimes \ket{\Omega\rangle} \, .
\end{equation} As introduced in the main text, $\ket{0}$ denotes the vacuum of the CFT while $\ket{\Omega\rangle}$ is the many-body Fock space vacuum. $\mathcal{O}_{\rm Bkg}$ is the background charge ensuring neutrality in the CFT correlator. Combining Eq.~\ref{eq:LLLOrbitalCylinder} and the mode expansion of the electronic operator on the cylinder: \begin{equation} \label{eq:ModeExpansionCylinder}
\mathcal{V} (w) = \sum_{j\in\mathbb{Z}}z^j \mathcal{V}_{-j} \quad , \quad \mathcal{V}_{-j} = \oint \frac{{\rm d}z}{2i\pi} z^{-j-1} \mathcal{V}(w) \, ,
\end{equation} we obtain
\begin{equation} \label{eq:FromIntegralToModes}
\exp\left[\int {\rm d}^2w \, \mathcal{V} (w) e^{-\frac{x^2}{2}} \otimes c^\dagger(w) \right] = \exp\left[\sum_{j \in \mathbb{Z}} \left(\dfrac{1}{\mathcal{N}_j} \mathcal{V}_{-j} \right) \otimes  c_j^\dagger \right] = \prod_{j\in\mathbb{Z}} \left[ \sum_{m_j \in \mathbb{Z}} \dfrac{1}{m_j!} \left( \dfrac{1}{\mathcal{N}_j} \mathcal{V}_{-j} \right)^{m_j} \otimes  \left( c_j^\dagger \right)^{m_j} \right] \, .
\end{equation} To derive the last equality, we have used the fact that all terms in the exponential commute. Indeed, the electronic modes $\mathcal{V}_{-j}$ have the same commutation relation as the bosonic creation operators because of the commutation properties of $\mathcal{V}(w)$. This equation allows for an explicit site-dependent MPS representation of the WF Eq.~\ref{eq:GenericMpsSecondQuant} over the occupation basis $\ket{\{m_j\}\rangle} = \prod_{j\in\mathbb{Z}} \frac{1}{\sqrt{m_j !}} \left( c_j^\dagger \right)^{m_j} \ket{\Omega\rangle}$:
\begin{equation} \label{eq:SiteDependentMPS}
\braket{\langle \{m_j\} }{\Psi \rangle} = \braOket{0}{ \mathcal{O}_{\rm Bkg} \, \left( \prod_{j\in\mathbb{Z}} A^{(m_j)}[j] \right) }{0} \quad \text{with: } \, A^{(m)}[j] = \dfrac{1}{\sqrt{m !}} \left( \dfrac{1}{\mathcal{N}_j} \mathcal{V}_{-j} \right)^m \, .
\end{equation} This is exactly the formula derived in Ref.~\cite{Regnault_ConstructionMPS}. Note that we did not make any assumption about geometry and Eq.~\ref{eq:SiteDependentMPS} is valid on the disk or sphere up to slight differences in the mode expansion of Eq.~\ref{eq:ModeExpansionCylinder} and the geometrical factors $\mathcal{N}_j$. What prevents us from obtaining a site independent representation is first the explicit dependence of our tensors Eq.~\ref{eq:SiteDependentMPS} on the geometrical factors $\mathcal{N}_j$ and second the accumulation of charge throughout the cylinder before the background charge, localized at one end, sets it to zero. Given the translation invariance on the cylinder geometry, those two problems may be solved and we here briefly summarize how to obtain an orbital-independent MPS description of the state.

We focus on the case where the underlying CFT consists of a neutral part and a U(1) part which carries the electric charge. To make the discussion more explicit, we consider the one-component bosonic Pfaffian state~\cite{MooreRead_CFTCorrelatorModelState}. Among the FQH model WFs, the Pfaffian state occupies a special place since it is the simplest and paradigmatic example of topologically ordered ground state hosting non-Abelian excitations~\cite{WilczekNayak_NonAbelianMR}. Its first quantized ground state WF reads: \begin{equation} \label{eq:MooreReadFirstQuantized}
\Psi_{\rm Pf} (z_1, \cdots , z_{N_b}) = {\rm Pf} \left(\dfrac{1}{z_i - z_j}\right) \prod_{i<j} (z_i-z_j) \, ,
\end{equation} where $N_b$ is the (even) number of bosons. It appears at filling fraction $\nu=1$ and can be exactly represented as a CFT correlator within our formalism. In that case the neutral CFT consists of a free chiral Majorana fermion $\psi$, and the corresponding electronic operator is \begin{equation} \label{eq:MooreReadElectronicOperatorIntro}
\mathcal{V}(z) = \psi (z) :e^{i\frac{1}{\sqrt{\nu}}\phi^c(z)}: \, ,
\end{equation} with $\phi^c$ the bosonic field associated to the electric charge. The Pfaffian factor is produced as the $N$-points correlation function of a free Majorana fermionic field $\psi$, while the Jastrow factor comes from the vertex operator (see Ref.~\cite{YellowBook}). Since Eq.~\ref{eq:MooreReadElectronicOperatorIntro} satisfies the correct commutation relation, we obtain the second quantized form of Eq.~\ref{eq:MooreReadFirstQuantized}:
\begin{equation} \label{eq:MooreReadCftCorrelator}
\ket{\Psi_{\rm Pf}\rangle} = \braOket{0}{ \mathcal{O}_{\rm Bkg} \, e^{\int {\rm d}^2 z \, \mathcal{V} (z) \otimes c^\dagger(z) } }{0} \otimes \ket{\Omega\rangle} \, .
\end{equation}  For a system of $N_b$ bosons in $N_{\rm orb}=N_b/\nu$ orbitals, the background charge reads $\mathcal{O}_{\rm Bkg}=\exp( -iN_b \sqrt{\nu^{-1}} \phi_0^c )=\exp( -i N_{\rm orb}\sqrt{\nu} \phi_0^c )$, $\phi_0^c$ is the bosonic zero-point momentum (see below). Spreading the background charge equally between the orbitals while accounting for the cylinder geometrical factors is detailed in Refs.~\cite{ZaletelMong_ExactMPS,Regnault_ConstructionMPS,Regnault_MPSHalperinStates}. Thanks to the relation \begin{equation} \label{eq:DefOfU}
U^{-1} A^{(m)}[j]U = A^{(m)}[j-1] \, , \quad 
U = \exp \left( -\gamma^2 L_0 - i\sqrt{\nu} \phi_0^c \right) \, ,
\end{equation} where $L_0$ denotes the total Virasoro zero-th mode, we can recast Eq.~\ref{eq:SiteDependentMPS} into the site independent MPS: 
\begin{equation}\label{eq:SiteIndependentMPS}
\braket{\langle \{m_j\} }{\Psi \rangle} = \braOket{0}{ \left( \prod_{j\in\mathbb{Z}} B^{(m_j)} \right) }{0} \quad \text{with: } \, B^{(m)} = \dfrac{1}{\sqrt{m !}} \left( \mathcal{V}_{0} \right)^m U \, .
\end{equation} In order to evaluate this MPS, we need to specify the CFT Hilbert space and to compute the action of $\mathcal{V}_0$ in this basis. We summarize the relevant notions to perform such calculation in the next section.

\subsection{B.2. Free Bosons and Majorana Fermions in CFT}

\subsubsection{B.2.1. Compactified Free Boson}

The free massless boson $\phi^c$ CFT has central charge $c=1$ (see for instance Ref.~\cite{YellowBook}). In this paper, we assume that it has compactification radius $R=\frac{1}{\sqrt{\nu}}$, with $R^2 \in \mathbb{N}$. Its two-point correlation function is given by $\langle \phi^c(z) \phi^c(w) \rangle =-\log (z-w)$ and its mode expansion on the plane reads:
\begin{equation} \label{eq:FreeBosonModeExpansion}
\phi^c (z) = \phi_0^c -i a_0 \log z + i \sum_{n\in \mathbb{Z}^*} \dfrac{1}{n} a_{n} z^{-n} .
\end{equation} The $a_n$ satisfy a U(1) Kac-Moody algebra: $[a_n,a_m]=n \delta_{m+n,0}$. This U(1)-symmetry implies the conservation of the current $J(z)=i R \partial \phi^c (z)$ and the U(1)-charge is measured by the zero-mode $R a_0$. The zero point momentum $\phi_0^c$ is the canonical conjugate of $a_0$, \textit{i.e.} $[\phi_0^c,a_0]=i$. As such, the operator $e^{i \sqrt{\nu} \phi_0^c}$ shifts the U(1) charge by one. Primary fields with respect to the U(1) Kac-Moody algebra are vertex operators of quantized charges: \begin{equation} \label{eq:PrimaryKacMoodyU1} \mathbf{V}_N (z) = : \exp \left(i \dfrac{N}{R} \varphi (z) \right):  \, , \end{equation} where $N \in \mathbb{Z}$ (resp. $N \in \mathbb{Z}+1/2$) for periodic (resp. anti-periodic) boundary condition on the cylinder. In particular $\mathbf{V}_{R^2}$ is a primary field of the theory, it will appear in all the electronic operators. The energy momentum tensor is $T^b(z) = -(1/2) :(\partial\phi^c)^2 :$ and its zero-th mode $L_0^b$ grades the states with respect to their conformal dimensions. Finally, a basis spanning the whole Hilbert space is labeled by a bosonic partition $\mu$ and the charge index $N$: \begin{equation} \ket{N, \mu} = \dfrac{1}{\sqrt{\Xi_\mu}} \prod_{i=1}^{\ell (\mu)} a_{-\mu_i} \ket{N} \, , \end{equation} where $\ell (\mu)$ is the length of the partition $\mu$ (\textit{i.e.} the number of non-zero elements), and the prefactor reads $\Xi_\mu = \prod_i i^{n_i} n_i!$ where $n_i$ is the multiplicity of the occupied mode $i$ in the partition $\mu$. They have conformal dimension: $L_0^b \ket{N, \mu} = (N^2/2+|\mu|)\ket{N, \mu}$, with $|\mu |=\sum_i \mu_i$. Evaluation of the vertex operators in this basis leads to the matrix elements given, for instance, in Ref.~\cite{Regnault_MPSHalperinStates}.

\subsubsection{B.2.2. Free Majorana Fermion}

The free Majorana $\psi$ CFT has central charge $c=1/2$ (see Ref.~\cite{Ginspard_CFTlectures} for a review) with respect to the stress energy tensor $T^\psi(z) = (1/2) : \psi \partial \psi:$. This neutral CFT divides in two pieces: one containing the vacuum state, and another obtained by acting with a twist operator $\sigma$ of dimension $1/16$ on the former. The mode expansion on the cylinder reads
\begin{equation}
\psi (z) = \sum_n \psi_{-n} z^n \, ,
\end{equation} where $n \in \mathbb{Z}+1/2$ in the untwisted sector and $n\in \mathbb{Z}$ in the twisted sector. The fermionic modes satisfy the anti-commutation relations $\{ \psi_n , \psi_m \} = \delta_{m+n,0}$.  A basis for the fermionic CFT Hilbert space is obtained by  repeated action of the modes $\psi_{-n}$ with $n>0$. We may label the states with a fermionic partition $\eta$ and an index $\bm{\tau} \in \{ \bm{1}, \bm{\sigma} \}$ to distinguish the untwisted and twisted sector:
\begin{equation}
\ket{\bm{\tau},\eta} = \prod_{i=1}^{\ell (\eta)} \psi_{-\eta_i} \ket{\bm{\tau}} \,  .
\end{equation} They have conformal dimension: $L_0^\psi \ket{\bm{\tau},\eta} =  (\varepsilon/16+|\eta|)\ket{\bm{\tau},\eta}$, with $\varepsilon=0$ in the untwisted sector and $1$ otherwise. The evaluation of $\psi (z)$ in this basis can be performed with the fermionic modes anti-commutation relations.

\subsection{B.3. Compatible MPS for the Halperin 220 and Pfaffian States}
\subsubsection{B.3.1. CFT and Projective Construction}

Numerical studies have consistently reported~\cite{Jolicoeur_SpinlessBosons,Ueda_PhaseDiagBilayerBoson,Regnault_EmergentPHsymmetryBilayer,Regnault_BilayerBosonicPhaseDiagram} that the Pfaffian phase may be stabilized in a bilayer FQH system of bosons at large inter-layer tunneling. More precisely, it is seen that the system undergoes a topological phase transition from the Halperin 220 phase to the bosonic Pfaffian phase. Motivated by these physical insights, we now show how to obtain an iMPS for the 220 Halperin state and the previously studied Pfaffian state within the same physical and auxiliary space. The conformal field description of the Halperin states usually relies on a charge $\phi^c$ and a spin $\phi^s$ chiral bosonic degree of freedom~\cite{Hansson_ReviewCFTforFQHE}. They essentially represent the spinful Luttinger liquid living at the edge of the Hall droplet, where spin-charge separation is expected~\cite{Voit_SpinCharge,Giamarchi_1Dphysics}. Starting from the $\mathbf{K}$-matrix~\cite{WenZee_Kmatrix}, it is possible to define two electronic operators (one for each spin component) as~\cite{Regnault_MPSHalperinStates}:
\begin{equation} \label{eq:VertexOperatorsSpinCharge}
\mathcal{V}^\uparrow (z) = : e^{i\phi^s(z)} : \, : e^{i\phi^c(z)} : \, , \quad  \mathcal{V}^\uparrow (z) = : e^{-i\phi^s(z)} : \, : e^{i\phi^c(z)} : \, ,
\end{equation} where the first (resp. second) component is relative to spin up (resp. down) particles. Since these electronic operators commute, and their OPEs only involve regular terms $\mathcal{V}^\tau(w_1)\mathcal{V}^\sigma(w_2) \sim_{\rm \scriptscriptstyle OPE} \mathcal{O}(1)$, the previous derivation can be straightforwardly reproduced for the Halperin state:
\begin{equation}
\ket{\Psi_{220}\rangle} = \braOket{0}{ \mathcal{O}_{\rm Bkg} \, e^{\int {\rm d}^2 z \left[ \mathcal{V}^\uparrow (z) \otimes c_\uparrow^\dagger(z) + \mathcal{V}^\downarrow (z) \otimes c_\downarrow^\dagger(z) \right]} }{0} \otimes \ket{\Omega\rangle} \, .
\end{equation} Specifying the number of particles in each spin components and their positions, and using the vertex operators OPE~\cite{YellowBook,Ginspard_CFTlectures}, the more familiar first quantized form is retrieved (see main text).

As previously stated, we are interested in adding tunneling between layers. As in the main text, we introduce the symmetric and anti-symmetric combinations $c_{e/o} (z) = (c_\uparrow(z)\pm c_\downarrow(z))/\sqrt{2}$. In this rotated basis, the Halperin 220 state becomes 
\begin{equation} \label{eq:HalperinSigmaXBasis}
\ket{\Psi_{220}\rangle} = \braOket{0}{ \mathcal{O}_{\rm Bkg} \, e^{\int {\rm d}^2 z \left[ \mathcal{V}^e (z) \otimes c_e^\dagger(z) + \mathcal{V}^o (z) \otimes c_o^\dagger(z) \right]} }{0} \otimes \ket{\Omega\rangle} \, ,
\end{equation} where we have defined $\mathcal{V}^{e/o} = (\mathcal{V}^\uparrow \pm \mathcal{V}^\downarrow)/\sqrt{2}$. In this language it is natural to introduce the real and imaginary part of a Dirac fermionic fields $\psi^R$ and $\psi^I$ in order to fermionize the spin vertex operators of Eq.~\ref{eq:VertexOperatorsSpinCharge} as $:e^{\pm i\phi^s}: = (\psi^R\pm i\psi^I)/\sqrt{2}$. Indeed, it allows to decouple the spin parts of the vertex operators used in Eq.~\ref{eq:HalperinSigmaXBasis}: 
\begin{equation} \label{eq:VertexOperatorsSigmaX}
\mathcal{V}^e (z) = \psi^R(z) \cdot \mathbf{V}(z) \, , \quad  \mathcal{V}^o (z) = i\psi^I(z) \cdot \mathbf{V}(z) \, ,
\end{equation} in which $\mathbf{V}=:e^{i\phi^c}:$ is a short hand notation for the vertex operator of the charge bosonic field. In the following, we choose to work within the CFT of a free boson coupled to that of a Dirac fermion. Such a choice allows for an exact description of the Halperin 220 state, with corresponding iMPS matrices:
\begin{equation}\label{eq:iMPSmatrices220}
B_{\rm H}^{(m_+,m_-)} = \dfrac{1}{\sqrt{m_+ ! m_-!}} \left( \mathcal{V}_0^e \right)^{m_+} \left( \mathcal{V}_0^o \right)^{m_-} U \, , \end{equation} where $U$ has been defined in Eq.~\ref{eq:DefOfU}. Tunneling between layers act in the rotated basis as a chemical potential favoring the symmetric combination $e$. Heuristically, it projects the state Eq.~\ref{eq:HalperinSigmaXBasis} into \begin{equation} 
\ket{\Psi_{\rm Pf}\rangle} = \braOket{0}{ \mathcal{O}_{\rm Bkg} \, e^{\int {\rm d}^2 z \, \mathcal{V}^+ (z) \otimes c_+^\dagger(z) } }{0} \otimes \ket{\Omega\rangle} \, ,
\end{equation} which is the exact representation of the Pfaffian state given in Eq.~\ref{eq:MooreReadCftCorrelator}. Performing the projection on the physical indices of the iMPS matrices Eq.~\ref{eq:iMPSmatrices220} leads to an MPS describing exactly the bosonic Pfaffian state on the cylinder: \begin{equation}
B_{\rm Pf}^{(m_+)} = B_{\rm H}^{(m_+,0)} \, .
\end{equation} The same construction applies equally well to any variational parameter $\theta$ (see main text) and $B_\theta^{(m_+,m_-)} = (\cos\theta)^{m_+} (\sin\theta)^{m_-} B_{\rm H}^{(m_+,m_-)}$.

\subsubsection{B.3.2. Auxiliary Space}

The CFT Hilbert space studied for the one-component Pfaffian state was composed of a single Majorana fermion and a compact boson. We need to enlarge this space to include the $\psi^I$ degree of freedom, over which $B_{\rm Pf}^{(m_+)}$ acts trivially. This is the redundancy in the auxiliary space mentioned in the main text, the iMPS describing the Pfaffian state in not injective as multiple boundary conditions lead to the same physical state. The only subtlety here will be in the treatment of the fermionic zero modes. Indeed, the two Majorana fields $\psi^R$ and $\psi^I$ are part of the same Dirac fermion $\Psi = (\psi^R+i \psi^I)/\sqrt{2}$ and hence share the same boundary condition. More precisely, the Hilbert space associated with the Dirac fermionic CFT also splits into two part: the untwisted sector $\bm{1}$ containing the vacuum where $\Psi$ has antiperiodic boundary condition when it winds around the cylinder, and a twisted sector $\bm{\sigma}$ corresponding to periodic boundary condition. A basis for the fermionic CFT Hilbert space is obtained by acting with $\psi_{-n}^R$ and $\psi_{-n}^I$ with $n>0$ on the highest weight states. For instance in the vacuum sector, we may label the states with two fermionic partitions $\eta$ and $\nu$:
\begin{equation}
\ket{\mathbb{I},\eta,\nu} = \prod_{i=1}^{\ell (\eta)} \psi_{-\eta_i}^R \prod_{i=1}^{\ell (\nu)} \psi_{-\nu_i}^I \ket{0} \,  , \quad \eta_i , \nu_i \in \mathbb{N}+1/2 \, .
\end{equation} 
In the twisted sector, the only differences are the mode index and the state on which the modes $\psi_{-n}^R$ and $\psi_{-n}^I$ act.  A subtlety arises in the treatment of the later due to the presence of zero modes. Because $\psi_0^R$ and $\psi_0^I$ anticommute, the dimension of the highest weigth vector space must be at least two in the twisted sector. To numerically implement these anticommutation relation, we introduce two degenerate twist fields  $\ket{\sigma_+}$ and $\ket{\sigma_-}$ of conformal dimension 1/8 spanning this vector space, over which $\psi_0^R$ and $\psi_0^I$ act as two Pauli matrices~\cite{Ginspard_CFTlectures}. The Hilbert space is then spanned by \begin{equation}
\ket{\sigma_\pm,\eta,\nu} = \prod_{i=1}^{\ell (\eta)} \psi_{-\eta_i}^R \prod_{i=1}^{\ell (\nu)} \psi_{-\nu_i}^I \ket{\sigma_\pm}  \,  , \quad \eta_i , \nu_i \in \mathbb{N}^* \, .
\end{equation}  

Assuming periodic boundary conditions for the electronic operator Eq.~\ref{eq:MooreReadElectronicOperatorIntro} introduces constraints between the sectors of the fermionic Hilbert space and the U(1)-charges of the boson. More precisely, the vertex operator and the fermionic fields should either be both periodic ($N \in \mathbb{Z}$ and twisted sector $\bm{\sigma}$), or both anti-periodic ($N \in \mathbb{Z}+1/2$ and untwisted sector $\bm{1}$). A general state of the Hilbert space thus reads: \begin{equation}
\ket{N,\mu} \otimes \ket{\bm{\tau}, \eta, \nu} 
\end{equation} with the previous constraint fulfilled. The action of the electronic operator's integer modes (Eq.~\ref{eq:ModeExpansionCylinder}) in this basis consists of a phase factor, coming from the fermionic operator, and of the non-trivial coefficient coming from the vertex operator~\cite{Regnault_ConstructionMPS,Regnault_MPSHalperinStates}. Altogether, it allows for the computation of the matrix elements of $\mathcal{V}_0^{e/o}$ while the action of $L_0$ makes the elements of $U$ explicit.

\section{C. Applying a Microscopic Hamiltonian without Matrix Product Operators}

Not only does Eq.~\ref{eq:SiteIndependentMPS} describes the LLL model states, but varying the boundary index of the MPS $\bra{\alpha}$ in the CFT Hilbert space also produces all quasihole zero-energy excitations of the system~\cite{Regnault_MPSnonAbelianQH}: \begin{equation}
\ket{\Psi_{\bra{\alpha}}\rangle } = \braOket{\alpha}{\prod_{j\in \mathbb{Z}} \left(e^{\mathcal{V}_0 \otimes c_j^\dagger} \,  U\right) }{0} \otimes \ket{\Omega \rangle} \, .
\end{equation} The physical relevance of the model WFs such as Eq.~\ref{eq:GenericMpsSecondQuant} was considerably substantiated by the finding of interacting Hamiltonians for which they are exact ground state. For instance, the bosonic Laughlin WF is the densest ground state for contact interactions in the LLL at filling $\nu=1/2$~\cite{Kivelson_PseudoPotentials}. The Pfaffian state completely screens a three-body interaction penalizing the clustering of three particles at the same position~\cite{HaldaneBernevig_ClusteringCondition}. In this section, we show how to apply microscopic Hamiltonian projected onto a single Landau level in the subspace spanned by  $\{ \ket{\Psi_{\bra{\alpha}}\rangle } \}$ without relying on Matrix Product Operators (MPOs).

\subsection{C.1. Mean Occupation Number}

\begin{figure}
	\centering
	\includegraphics[width=0.7\columnwidth]{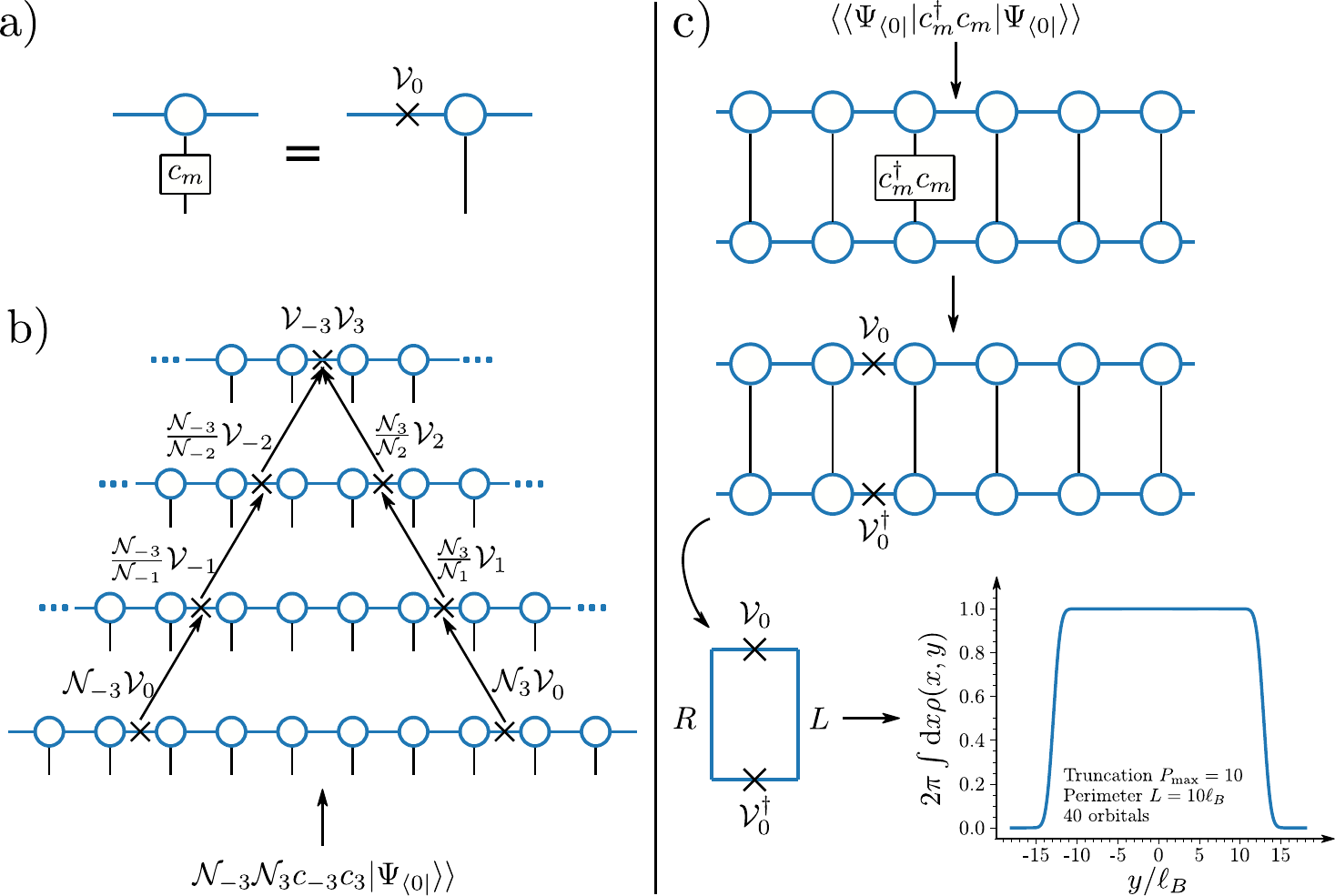}
	\caption{\emph{a) Due to the structure of the iMPS ansatz Eq.~\ref{eq:SiteIndependentMPS}, the action of annihilation operators may be computed as insertion of a $\mathcal{V}_0$ operator at the right position in the chain. We depict how it works for the mean occupation number (see text), from which we can infer the density profile of the bosonic Pfaffian states over 40 orbitals and a cylinder perimeter $L=10 \ell_B$. We reach convergence for a truncation parameter $P_{\rm max}=10$. b) For long range Hamiltonians, we may furthermore use the explicit form of Eq.~\ref{eq:DefOfU} to gather all insertions at a single location in the chain. This technique allows a highly efficient numerical computation of the interaction energy, even though the Hamiltonian is long ranged in the LLL basis.}}
	\label{Fig:DensityComputationMPO}
\end{figure}

To exemplify the construction, let us consider the simple example of the orbital occupation number $n_m = c_m^\dagger c_m$, $m\in\mathbb{Z}$, from which the density may be read off by mapping the LLL to real space Eq.~\ref{eq:LLLOrbitalCylinder}. Computing its expectation value on $\ket{\Psi_{\bra{\alpha}}\rangle }$ amounts to the computation of the overlap of $c_m \ket{\Psi_{\bra{\alpha}}\rangle }$ with itself. Noticing that \begin{equation} \label{eq:MicroscopicActionAnnihilation}
c_m \ket{\Psi_{\bra{\alpha}}\rangle } = \braOket{\alpha}{\prod_{j < m} \left(e^{\mathcal{V}_0 \otimes c_j^\dagger} \,  U\right) \mathcal{V}_0 \prod_{j \geq m} \left(e^{\mathcal{V}_0 \otimes c_j^\dagger} \,  U\right) }{0} \otimes \ket{\Omega \rangle} \, , 
\end{equation}, which is graphically depicted in Fig.~\ref{Fig:DensityComputationMPO}a, this overlap may be computed in the MPS transfer matrix formalism straightforwardly. To derive this identity, we used the fact that both the vertex operator modes and the creation/annihilation operators share the same commutation relations. Denoting $R$ and $L$ the right and left eigenvectors of the iMPS transfer matrix, represented as square matrix of size the truncated CFT Hilbert space dimension, the graphical representation of Fig.~\ref{Fig:DensityComputationMPO}c formally reads: \begin{equation} \label{eq:MicroscopicActionDensity}
\bra{\langle \Psi_{\bra{\alpha}} } c_m^\dagger c_m \ket{\Psi_{\bra{\alpha}}\rangle } = \dfrac{\left( L ; \mathcal{V}_0 R \mathcal{V}_0^T \right)}{\left(L ; R \right)} \, ,
\end{equation} where we have used the scalar product $\left( A ; B \right) = {\rm Tr}\, (A^\dagger B)$. We can summarize this result as follows. Imposing the same commutation relations for the vertex operators and the creation/annihilation operators, and exploiting the exponential form of our MPS (Eq.~\ref{eq:SiteIndependentMPS}) we can translate an operator acting on the physical degrees of freedom of our MPS onto an insertion of operators on the auxiliary space. In the following, we show how to generalize this to more intricate microscopic operators (such as model Hamiltonians) acting in the zero-energy quasiholes subspace $\{ \ket{\Psi_{\bra{\alpha}}\rangle } \}$. 

\subsection{C.2. Delta Interactions}

~\cite{Cirac_CalculusContinuousMPS}The interacting Hamiltonians stabilizing FQHE phases are notoriously hard to implement as MPO~\cite{Zaletel_DMRG_Hall,Zaletel_DMRG_Multicomponent}, be it because of their long range expression in the LLL occupation basis~\cite{Kivelson_PseudoPotentials} or the multi-body terms involved~\cite{MooreRead_CFTCorrelatorModelState,HaldaneBernevig_ClusteringCondition}. We show here how to use the previous prescription for delta interaction that we will use in the article, the derivation may be repeated for the three-body interactions stabilizing the Pfaffian state Eq.~\ref{eq:MooreReadFirstQuantized} as well. Once projected to the LLL, contact interaction may be written down as~\cite{Thomale_PseudoPotentials}: \begin{equation} \label{eq:PseudopotDeltaLLL}
H_{\rm int} = \frac{\sqrt{2\pi}}{L} \sum_{j \in \mathbb{Z}} \left( Q_j^\dagger Q_j +  Q_{j+1/2}^\dagger Q_{j+1/2} \right) \quad  \text{where: }  Q_j = \sum_\ell e^{(\gamma \ell)^2} c_{j-\ell}c_{j+\ell} \, .
\end{equation} In the definition of $Q_j$, the index $\ell$ runs over all integers (resp. half-integers) if $j \in \mathbb{Z}$ (resp. $j\in \mathbb{Z}+1/2$). The sum over all centers $j$ in Eq.~\ref{eq:PseudopotDeltaLLL} encodes the translation invariance along the cylinder axis. Let us thus focus on the terms that arise for $j=0$ and compute $\bra{\langle \Psi_{\bra{\alpha}} } Q_0^\dagger Q_0 + Q_{1/2}^\dagger Q_{1/2} \ket{\Psi_{\bra{\alpha}}\rangle }$. Using the argument of the previous discussion, this expectation value may be computed within the transfer matrix framework as a sum of terms with two $\mathcal{V}_0$ and two $\mathcal{V}_0^\dagger$ insertions at different position in the chain. It may seem hopless to compute so many coefficients exactly given the arbitrary large distance between insertions (the same difficulty arises for an MPO description of this interacting Hamiltonian, see Refs.~\cite{Zaletel_DMRG_Hall,Zaletel_DMRG_Multicomponent}). However, we can rely on the explicit form Eq.~\ref{eq:DefOfU} to \emph{exactly} shifts the location of the insertions along the chain by shifting the mode index of the matrix and accounting for the geometrical factors.

Let us first show how the derivation goes for one term appearing when expanding $Q_0^\dagger Q_0$. We take $m,n \in \mathbb{Z}$ and would like to compute $\bra{\langle \Psi_{\bra{\alpha}} } \mathcal{N}_{-m} c_{-m}^\dagger   \mathcal{N}_{m} c_{m}^\dagger  \mathcal{N}_{-n} c_{-n} \mathcal{N}_{n} c_{n}  \ket{\Psi_{\bra{\alpha}} \rangle }$. We insert, as before, two $\mathcal{V}_0$ at positions $\pm n$ and two $\mathcal{V}_0^\dagger$ at $\pm m$. As depicted in Fig.~\ref{Fig:DensityComputationMPO}b, we repeatedly use the commutation relation with $U$ to show that the expression is equivalent to the insertion of $\mathcal{V}_{-n} \mathcal{V}_{n}$ and $\mathcal{V}_{-m}^\dagger \mathcal{V}_{m}^\dagger$ both at position zero. Taking the sum over $m$ and $n$, we have:
\begin{equation} 
\bra{\langle \Psi_{\bra{\alpha}} } Q_0^\dagger Q_0 \ket{\Psi_{\bra{\alpha}}\rangle } = \dfrac{\left( L ; \mathcal{Q}_0 R \mathcal{Q}_0^T \right)}{\left(L ; R \right)} \quad \text{with: } \mathcal{Q}_0 = \sum_{\ell \in \mathbb{Z}} \mathcal{V}_{-\ell} \mathcal{V}_{\ell} \, .
\end{equation} We can make the exact same derivation for $Q_{1/2}^\dagger Q_{1/2}$, the only difference is the geometrical factors that do not completely cancel out: \begin{equation}
\mathcal{Q}_{1/2} = e^{3\gamma^2/4} \sum_{\ell \in \mathbb{Z}} \mathcal{V}_{-\ell+1} \mathcal{V}_{\ell} \, .
\end{equation} Using the translational invariance along the cylinder, we found a way to compute the energy per orbital of the system in terms of insertions of local tensors. This method is numerically efficient as it only requires to compute the matrix elements of the electronic operators modes, which is as difficult as building the MPS matrices themselves and scales linearly with the bond dimension of the MPS.  We have checked that the results \emph{match} with exact diagonalization in finite size for the contact interaction and multiple FQH model states, be it the Laughlin state (the exact densest ground state of this interaction) or the Pfaffian state (only a fairly good description of the ground state at $\nu=1$ for this interaction).

\section{D. Anyon-Anyon Correlation Lengths}

\begin{figure}
	\centering
	\includegraphics[width=0.6\columnwidth]{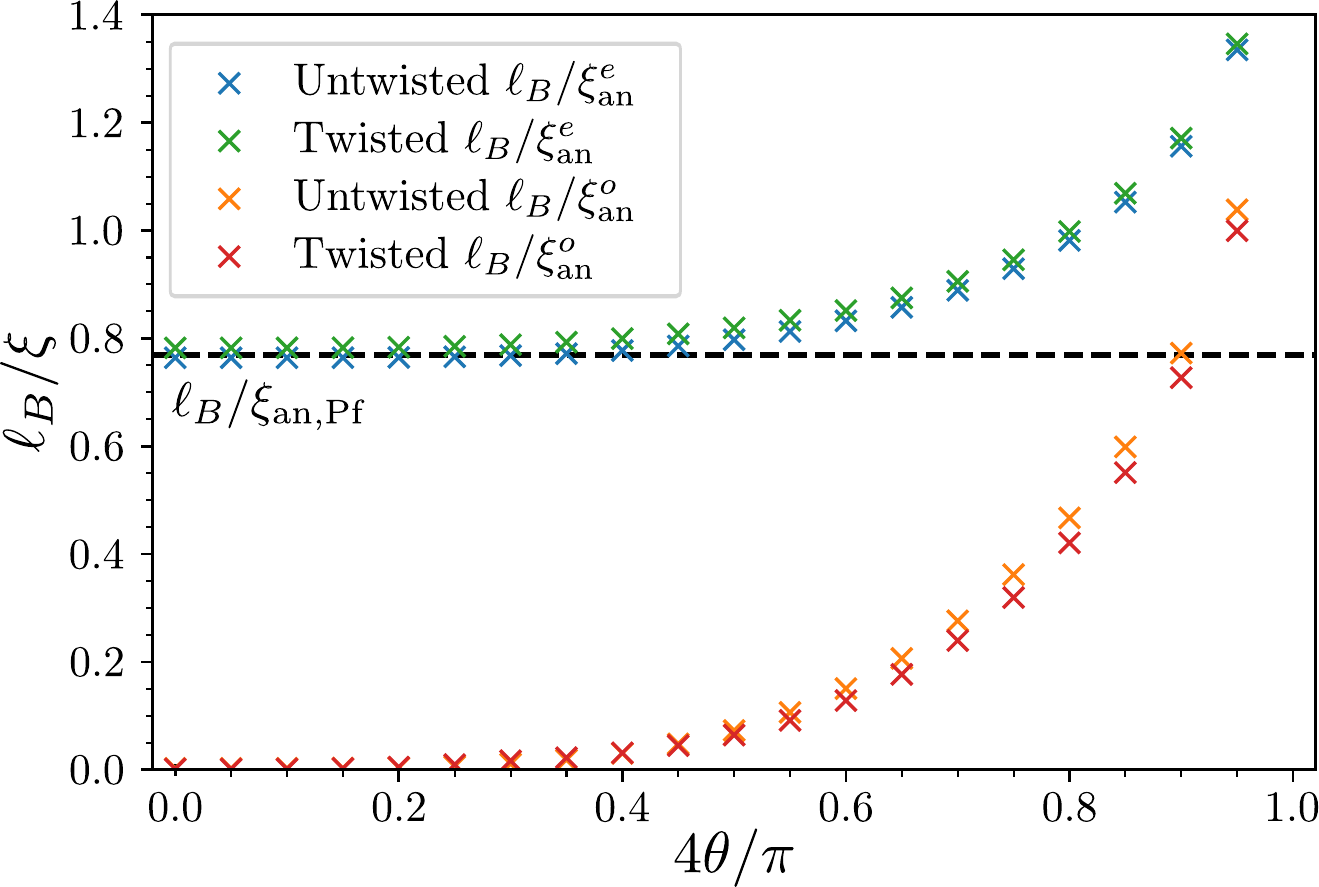}
	\caption{Inverse of the even and odd anyonic correlation lengths as a function of the variational parameter $\theta$. We observe the same features as in the main text for the decay of local observable correlation lengths.  }
	\label{Fig:AnyonicCorrelLength}
\end{figure}

The bosonic operators $\mathcal{V}^{e/o}$ change both $N_f$, the number of $\Psi$ fermions, and $N$ the U(1)-charge by one unit. The four topological sectors introduced in the main text are thus defined by $\tau \in \{\bm{1}, \bm{\sigma} \}$ and the parity of $N + N_f$. The background charge $U$ also leaves these sectors invariant every two orbitals. The MPS transfer matrix over two orbitals~\cite{Regnault_ConstructionMPS} \begin{equation}
	\mathbb{T} = \left( \sum_{m^+,m^-} B_\theta^{(m_+,m_-)} \otimes \left(B_\theta^{(m_+,m_-)}\right)^* \right)^2
\end{equation} is hence block diagonal $\mathbb{T} = \sum_{a,b} \mathbb{T}_{a,b}$ in the topological sectors defined above and labeled by $a$ and $b$. The four physical states built on the infinite cylinder are related to the leading right and left eigenvectors, which are located in the diagonal sectors $\mathbb{T}_{a,a}$. Since local observables only involve insertion of $\mathcal{V}^{e/o}$, the correlation functions of such observables only involve transfer matrix excited state lying in the same block. Thus the correlation lengths $\xi_{\rm loc}^e$ and $\xi_{\rm loc}^o$ are numerically extracted by diagonalization of the diagonal sectors $\mathbb{T}_{a,a}$. 

The excited states of the transfer matrix coming from the other block $\mathbb{T}_{a,b}$, with $a \neq b$, are only dialed when considering anyonic excitations. For instance, $\Psi$ and $:e^{i\phi^c}:$ change either $N_f$ or $N$ by one and thus shift the topological sector. The correlation function corresponding to insertions of these anyonic excitations at different positions also decays exponentially, with different correlation lengths $\xi_{\rm an}^e$ (resp. $\xi_{\rm an}^o$) related this time to the first even (resp. odd) eigenstate in the off-diagonal blocks of the tansfer matrix. While the precise values of $\xi_{\rm an}^e$ and $\xi_{\rm an}^o$ differ from $\xi_{\rm loc}^e$ and $\xi_{\rm loc}^o$, we observe the exact same features in both cases, as depicted in Fig.~\ref{Fig:AnyonicCorrelLength}. Even anyonic excitations are well screened all along the transition, while $\xi_{\rm loc}^o$ diverges when $\theta \to 0$ where antisymmetric anyonic excitations become transparent to the low-energy degrees of freedom of the system (see main text). 

\section{E. Extraction of $c$ : Numerical Details}

\begin{figure}
	\centering
	\includegraphics[width=0.8\columnwidth]{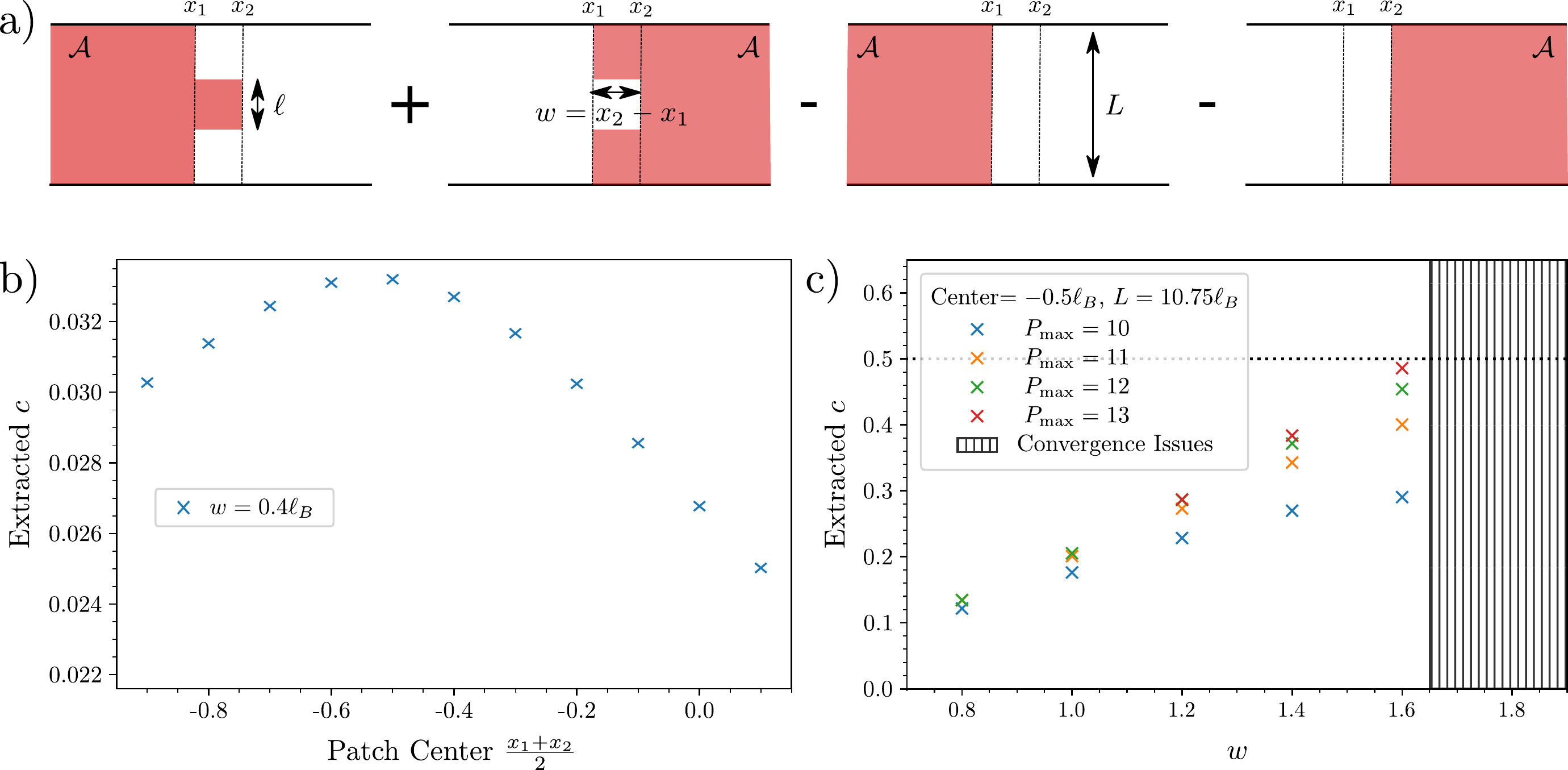}
	\caption{a) Addition-subtraction scheme removing area law contributions at $x_1$ and $x_2$ to the EE.  It allows to focus on the critical term Eq.~\ref{eq:CriticalContribution}. The top and bottom lines of each image should be identified to form the cylinder geometry. b) Extraction of the central charge $c$ as a function of the patch center. The largest critical contribution arises when the patch is at the interface mode exact location, here for $\frac{1}{2}(x_1+x_2) = -0.5\ell_B$. Using a small patch width $w=0.4 \ell_B$ allows to obtain converged results at a moderate bond dimension truncation parameter, $P_{\rm max}=11$, easing the numerical computations. c) With large widths $w$, the patch covers more of the interface critical mode and the extracted central charge $c$ increases. Compared to Refs.~\cite{OurNatComms_H2L2,OurNatComms_H3L3}, we are plagued by larger bulk correlation lengths and consequently more important area law contributions. They prevent us to see a clear plateau of $c$ when $w$ grows, before the saturation of the MPS finite bond dimension (controlled by the truncation parameter $P_{\rm max}$).}
	\label{Fig:MethodLevinWen}
\end{figure}

\subsection{E.1. Summary of the Method}

To characterize the one-dimensional interface effective theory, we compute the Real Space Entanglement Spectrum (RSES) for a bipartition for which the part $\mathcal{A}$ consists of a rectangular patch of length $\ell \in [0,L]$ along the compact dimension and width $w$ along the $x$-axis, \textit{i.e.} we break the rotational symmetry along the cylinder perimeter. To fully harness the power of the iMPS approach, it is convenient to add a half infinite cylinder to the rectangular patch (see Fig.~\ref{Fig:MethodLevinWen}a). The way to perform such a calculation with the exact MPS of FQH model WFs written as CFT correlators is fully explained in Ref.~\cite{OurNatComms_H3L3}. It was argued in Ref.~\cite{OurNatComms_H2L2} that the addition-subtraction scheme depicted on Fig.~\ref{Fig:MethodLevinWen}a removes the area law and the corner contributions that may arise. Hence, up to a constant $f(w)$ which depends on $w$, we expect the resulting Entanglement Entropy (EE) $S(\ell,w)$ to have the following form: \begin{equation} \label{eq:CriticalContribution}
S(\ell,w) = 2 \dfrac{c}{6} \log \left[\sin\left(\dfrac{\pi \ell}{L}\right)\right] + f(w) \, ,
\end{equation} where $c$ is the chiral anomaly of the interface effective theory. By considering either $S(\ell,w)-S(L/2,w)$ or the derivative $\partial_\ell S(\ell,w)$ (see Fig.~\ref{Fig:ResultsCExtract}a), we get rid off the constant term $f(w)$ and can perform a one-parameter fit of the numerical data to extract $c$. 

\subsection{E.2. Fixing the Parameters}

The addition-subtraction scheme only cancels the area law contributions in the final result, but they do appear in all the numerically computed EE. If the perimeter or the patch width is too large, the EE of some contributions of the addition-subtraction scheme will saturate the finite MPS bond dimension, limiting the extraction of the critical contribution Eq.~\ref{eq:CriticalContribution}. To consider small widths $w$, we should center the patch at the precise location of the interface edge mode. We can numerically determine its position by shifting the patch and observe where the largest critical contribution appears. We varied the center of the patch with a reasonable truncation parameter $P_{\rm max}=11$ and a small patch width $w=0.4 \ell_B$ (this combination allows to obtain converged results with short computation times, see Fig.~\ref{Fig:MethodLevinWen}c), and extracted the central charge. The results depicted in Fig.~\ref{Fig:MethodLevinWen}b indicate that the interface critical mode is located slightly in the Halperin 220 phase, at $x=-0.5\ell_B$. We fix the center of the patch at this position thereafter.

We then increase the width $w$ until the patch covers entirely the interface critical mode and hopefully reaching a plateau for $c$, as in Refs.~\cite{OurNatComms_H2L2,OurNatComms_H3L3}. While the extracted central charge increases with $w$, so do the area law contributions to the EE. Obtaining a converged result for the subleading critical term Eq.~\ref{eq:CriticalContribution} thus requires increasingly large bond dimensions, see Fig.~\ref{Fig:MethodLevinWen}c. Roughly speaking, the area contributions at fixed truncation parameter $P_{\rm max}$ should not saturate the EE which fixes a upper bound for $L+2w$. There are however two requirements to fulfill. The perimeter should be much greater than the two bulk correlation lengths not to suffer from finite size effects, and the width large enough to cover the entire critical mode.

\subsection{E.3. Results}

\begin{figure}
	\centering
	\includegraphics[width=\columnwidth]{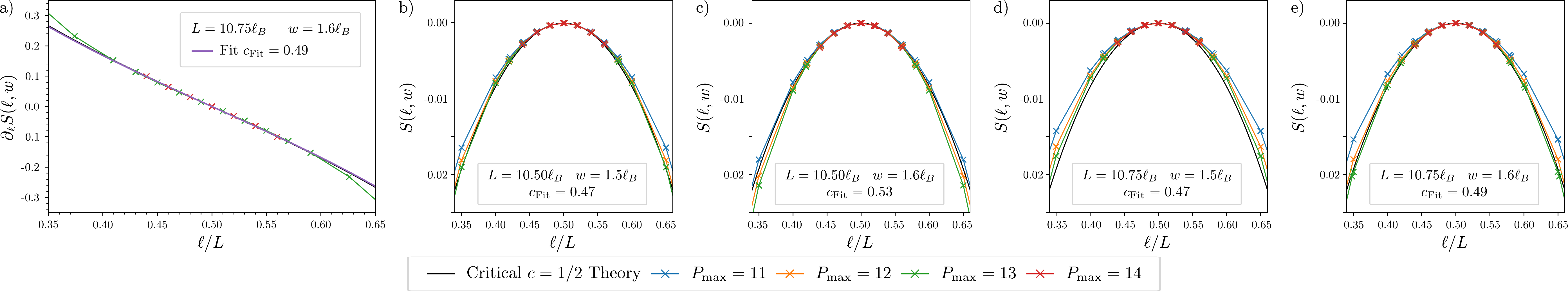}
	\caption{a) One-parameter fit of $\partial_\ell S(\ell,w)$ to numerically extract the central charge $c_{\rm Fit}$, here exemplified for $L=10.75\ell_B$ and $w=1.6\ell_B$. Taking the derivative allows to get rid off constant contributions to the EE and to isolate the one of the interface critical mode. b-e) Comparison of the numerically extracted $S(\ell,w)$ and Eq.~\ref{eq:CriticalContribution} in the parameter range $w \in \{1.5\ell_B, 1.6\ell_B \}$ and $L\in \{10.50 \ell_B, 10.75\ell_B\}$. In this parameter range, we do not suffer from finite size effects and our numerical results converge when increasing the bond dimension (measured by $P_{\rm max}$).}
	\label{Fig:ResultsCExtract}
\end{figure}

Plagued by larger bulk correlations lengths and greater linear coefficient for the area law terms compared to Refs.~\cite{OurNatComms_H2L2,OurNatComms_H3L3}, we were only able to find a small region in $L$ and $w$ where we reached convergence and where the extracted $c$ extracted did not depend small changes of the perimeter. In Fig.~\ref{Fig:ResultsCExtract}b-e, we show the convergence of $S(\ell,w)$ with respect to the truncation parameter $P_{\rm max}$ for two width $w \in \{1.5\ell_B, 1.6\ell_B \}$ and two perimeters $L=10.5\ell_B$ and $L=10.75\ell_B$. In all cases, the fitted central charge $c_{\rm Fit}$ agrees with the theoretical expectation for a chiral Majorana fermion $c=1/2$. The best result $c_{\rm Fit}=0.49$ is obtained for the largest perimeter and wider patch that we could safely consider at the reachable truncation parameters $P_{\rm max}$ (see Fig.~\ref{Fig:ResultsCExtract}).

\end{document}